\documentclass{article}
\usepackage{ltwol2e}
\usepackage{psfig}

\def\D0{D\O}
\def\bsg{$b\to s\gamma$}
\def\etal{{\it et al.}}
\def\prl{Phys. Rev. Lett.}

\begin{document}
%
%
\title{HEAVY QUARK PRODUCTION AND DECAY: $t$, $b$, AND ONIA$^*$}
\author{RICHARD PARTRIDGE}
\address{Brown University, Department of Physics, Providence, RI 02912,
USA\\E-mail: partridge@hep.brown.edu}   
%
%
\twocolumn[\maketitle\abstracts{ This paper summarizes a variety of
recent results on heavy quark production and decay. Considerable
progress has been made by CDF and \D0\ in measuring
top quark production and decay properties.
The top quark mass been measured with an uncertainty that is
less than 3\%, a relative precision that is better than has been
achieved for other quarks.
Measurements of the top production and decay properties are
consistent with the Standard Model predictions, but are
generally limited in precision by the small number
of top quark events in the present data samples.
ALEPH and CLEO have measured the branching ratio for the 
flavor changing neutral current
decay \bsg\ and find good agreement with the Standard Model.
Measurements of $b$-quark production cross sections in
$p\bar p$ collisions by CDF and \D0\ and in $e p$ collisions
by H1 continue to show discrepancies between theory and experiment.
CDF and DELPHI have measured the $b$-quark fragmentation fractions
with better accuracy than previous measurements.
Finally, measurements of charmonium production by \D0, E771, and L3
are consistent with predictions of the color octet model.
}]
\section{Introduction}
The topic of heavy quark production and decay covers quite a wide range
of material, exploring many different aspects of the Standard Model.\footnote[0]{
$^*$Plenary talk given at the XXIX International Conference on High Energy
Physics, Vancouver, Canada.}
To reduce the broad scope of the topic to a more manageable subset,
this paper focuses on top quark properties and the subset of bottom
and charm topics that were presented in the corresponding parallel
session of the conference.
Thus, the sections that follow describe the properties of the top quark,
measurements of the branching ratio for \bsg, $b$-quark
production in $p\bar p$ and $ep$ collisions, $b$-quark fragmentation
fractions, and measurements of charmonium production.
\section{Top Quark Properties}
In the three years that have elapsed since the
discovery~\cite{discovery1,discovery2}
of the top quark by CDF and \D0, a new field of inquiry has developed.
The goals of this field include making precise measurements of top quark
properties, probing features of the Standard Model that are unique to the
top quark, and searching for deviations from Standard Model predictions
that would indicate new physics.

To date, the greatest progress has been made in the measurement of
the top quark mass, which is interesting on a number of grounds.
As a fundamental parameter of the Standard Model, its value
contributes to the very definition of the Standard Model.
Once the top quark mass is fixed, the remaining top quark production
and decay properties can be predicted.
The top quark makes large contributions to various loop diagrams,
contributing to the $W$ boson propagator and the FCNC decay
\bsg.
Thus, measurement of the top quark mass, in conjunction with
measurements of the $W$ boson mass and electroweak parameters, allows the
mass of the Standard Model Higgs boson to be constrained and the
rate for \bsg\ to be predicted.
There may also be connections between the top quark mass and electroweak
symmetry breaking. For example, in supergravity inspired SUSY models,
radiative corrections provide a natural explanation for the 
``Mexican Hat'' shape of the Higgs potential provided the top quark
mass lies in the range 150--200 GeV/$c^2$.
The very fact that the top quark is so much heavier than the other
fermions in the Standard Model would seem to require an explanation
that can only come from new physics beyond the Standard Model.
The vastly different mass scales among the fundamental fermions has
been dramatically illustrated at the conference, where the observation
of neutrino oscillations by Super-Kamiokande~\cite{superk} appear to
indicate that $0<m_\nu<10^{-12}\ m_t$.

The large top quark mass is also responsible for many of the unique
features encountered in top quark physics:
\begin{itemize}
\item Since $m_t >> m_W$, real $W$ bosons are produced in top decay and can
be reconstructed from the decay products.
\item The Standard Model predicts $\Gamma_t >> \Lambda_{QCD}$, allowing the
study of a ``free'' quark that decays before it can hadronize.
\item The large top quark mass opens the possibility that new particles not
predicted by the Standard Model may be found in top quark decays.
\end{itemize}

While the top quark offers exciting opportunities to study new aspects of the
Standard Model, it will be difficult to shed new light on many of the
traditional topics in heavy quark physics.
For example, past discoveries of a new quark flavor have provided new
opportunities to improve our understanding of the CKM mixing matrix.
This is unlikely to be the case for the top quark discovery.
For a $3\times 3$ unitary CKM matrix, the CKM elements related to top quark
decay are tightly constrained by other measurements:~\cite{CKM}
\begin{equation}
\begin{array}{rcl}
|V_{tb}|^2 &=& 0.9985\pm 0.0002 \\
|V_{ts}|^2 &=& 0.0015\pm 0.0002 \\
|V_{td}|^2 &=& 0.00007\pm 0.00005
\end{array}
\label{eq:CKM}
\end{equation}
Given the difficulty in distinguishing $t\to s$ or $t\to d$ decays from the
dominant $t\to b$ decay mode, it is hard to see how studies of top quark
decays could further improve our understanding of the CKM matrix within
the context of the Standard Model. 
We can, however, test the Standard Model by measuring the single top
production cross section, which directly measures $|V_{tb}|^2$.
A non-unitary CKM
matrix or a fourth generation of quarks with substantial off-diagonal
CKM elements would cause $|V_{tb}|^2$ to deviate from the value given
in Eq.~(\ref{eq:CKM}).

In the sections below, the characteristics of top quark production,
measurements of the top quark mass, and various
tests of Standard Model predictions for top quark are described.
\subsection{Top Quark Production at the Tevatron}
At the Tevatron collider, top quark pair production occurs mostly through $q\bar q$
annihilation (90\%), with a small contribution from $gg$ diagrams (10\%). The
top quarks are expected to decay almost exclusively via $t\to Wb$, so the
final state will contain two $b$-quarks and the $W$ boson decay products.

The top quark decay channels are classified according to the $W$ boson decays.
The dilepton channel has both $W$ bosons decaying leptonically to $e\nu$ or
$\mu\nu$; it has a small (5\%) branching ratio but relatively little background.
The lepton~+~jets channel has one $W$ boson decaying leptonically and the
other hadronically; it has a large branching ratio (30\%) and a
substantial background from $W$~+~jets production.
The all-jets channel has both $W$ bosons decaying hadronically;
it has the largest branching ratio (44\%) and a very large QCD
multijet background.
The $\tau$ channels have one or both $W$ bosons decaying to $\tau\nu$;
they have a large aggregate branching ratio (21\%), but are difficult
to identify with the present detectors.

The lepton~+~jets channel is further divided according to the method used to
reduce the $W$~+~jets background:
\begin{itemize}
\item The SVX $b$-tag mode requires a displaced vertex in the CDF silicon
vertex detector consistent with $b$-quark decay.
\item The Lepton $b$-tag mode requires a non-isolated lepton consistent
with semileptonic $b$-quark decay.
\item The Topological mode places cuts on the kinematic variables $H_T$ and
Aplanarity.
\end{itemize}

Table~\ref{tab:events} shows that CDF and \D0\ observe a clear excess of
events over the expected background in the dilepton, lepton~+~jets,
and all-jets channels.
The two experiments have followed somewhat different strategies in defining
their event samples, with CDF taking advantage of their silicon vertex
detector to identify $b$-quarks, while \D0\ makes greater use of kinematic
variables to reduce backgrounds.
Despite considerable effort in optimizing these analyses, the present
sample of top candidates is small and most of the measurements
described below are dominated by statistical errors.
\begin{table}[hbt]
\begin{center}
\caption{The observed number of events and expected backgrounds for the top decay
channels studied by the CDF and \D0\ experiments. The $e\nu$ channel requires
a large transverse mass and is sensitive to $\tau$, dilepton, and lepton~+~jets
events that fail the standard cuts. The event selection criteria and background
estimation techniques are described in Refs.~5--8.}\label{tab:events}
\vspace{0.2cm}
\begin{tabular}{|l|c|c|c|c|} 
\hline
& \multicolumn{2}{|c|}{\raisebox{0pt}[12pt][6pt]{\D0}} &
\multicolumn{2}{|c|}{\raisebox{0pt}[12pt][6pt]{CDF}} \\
\cline{2-5}
\raisebox{0pt}[12pt][6pt]{Channel} & 
\raisebox{0pt}[12pt][6pt]{Evts} & 
\raisebox{0pt}[12pt][6pt]{BG} &
\raisebox{0pt}[12pt][6pt]{Evts} &
\raisebox{0pt}[12pt][6pt]{BG} \\
\hline
\raisebox{0pt}[12pt][6pt]{Dilepton} &
\raisebox{0pt}[12pt][6pt]{5} & 
\raisebox{0pt}[12pt][6pt]{$1.4\pm 0.4$} & 
\raisebox{0pt}[12pt][6pt]{9} & 
\raisebox{0pt}[12pt][6pt]{$2.4\pm 0.5$} \\
\hline
\raisebox{0pt}[12pt][6pt]{Lepton + Jets} &
\raisebox{0pt}[12pt][6pt]{-} & 
\raisebox{0pt}[12pt][6pt]{-} & 
\raisebox{0pt}[12pt][6pt]{34} & 
\raisebox{0pt}[12pt][6pt]{$9.2\pm 1.5$} \\
\raisebox{0pt}[12pt][6pt]{(SVX $b$-tag)} & & & & \\
\hline
\raisebox{0pt}[12pt][6pt]{Lepton + Jets} &
\raisebox{0pt}[12pt][6pt]{11} & 
\raisebox{0pt}[12pt][6pt]{$2.4\pm 0.5$} & 
\raisebox{0pt}[12pt][6pt]{40} & 
\raisebox{0pt}[12pt][6pt]{$22.6\pm 2.8$} \\
\raisebox{0pt}[12pt][6pt]{(Lepton $b$-tag)} & & & & \\
\hline
\raisebox{0pt}[12pt][6pt]{Lepton + Jets} &
\raisebox{0pt}[12pt][6pt]{19} & 
\raisebox{0pt}[12pt][6pt]{$8.7\pm 1.7$} & 
\raisebox{0pt}[12pt][6pt]{-} & 
\raisebox{0pt}[12pt][6pt]{-} \\
\raisebox{0pt}[12pt][6pt]{(Topological)} & & & & \\
\hline
\raisebox{0pt}[12pt][6pt]{All-jets} &
\raisebox{0pt}[12pt][6pt]{41} & 
\raisebox{0pt}[12pt][6pt]{$24.8\pm 2.4$} & 
\raisebox{0pt}[12pt][6pt]{187} & 
\raisebox{0pt}[12pt][6pt]{$142\pm 12$} \\
\hline
\raisebox{0pt}[12pt][6pt]{$e\tau$, $\mu\tau$} &
\raisebox{0pt}[12pt][6pt]{-} & 
\raisebox{0pt}[12pt][6pt]{-} & 
\raisebox{0pt}[12pt][6pt]{4} & 
\raisebox{0pt}[12pt][6pt]{$\approx 2$} \\
\hline
\raisebox{0pt}[12pt][6pt]{$e\nu$} &
\raisebox{0pt}[12pt][6pt]{4} & 
\raisebox{0pt}[12pt][6pt]{$1.2\pm 0.4$} & 
\raisebox{0pt}[12pt][6pt]{-} & 
\raisebox{0pt}[12pt][6pt]{-} \\
\hline
\end{tabular}
\end{center}
\end{table}
\vspace*{3pt}
\subsection{Measurement of the Top Quark Mass}
Measurements of the top quark mass have now been made in all of the
Standard Model decay channels: lepton~+~jets, dilepton, and all-jets.
These measurements are described in the sections below, followed by
a discussion of the mass results.  
\vspace{5pt}
\par\noindent
\underline{Lepton~+~Jets Channel}
\vspace{5pt}
\par\noindent
The most accurate measurement of the top quark mass is obtained
in the lepton~+~jets channel, where one of the top quarks decays
semileptonically (e.g., $t\to \ell \nu b$)  and the other
hadronically (e.g., $\bar t\to q\bar q \bar b$). 

In principle, the top quark mass can be measured
directly by calculating the invariant mass of the decay products from the
hadronically decaying top quark. However, it is not easy to
identify these decay products.
Initial State Radiation (ISR) introduces
additional partons that are not associated with top decay, while Final
State Radiation (FSR) introduces additional partons that should be
included among the top decay products.
Both CDF and \D0\ use the four highest $E_T$ jets in
reconstructing the top event, which Monte Carlo studies~\cite{d0ljetm}
indicate correctly identifies the partons from top decay 50\% of the time.

Even when the top quark decay products are correctly identified, there
are 12 different permutations for assigning jets to partons for events
with no $b$-tag, 6 permutations for events with a single $b$-tag, and
2 permutations when both $b$-quarks are tagged.
To improve the probability of picking the correct jet permutation, a
two constraint (2C) kinematic fit is performed that constrains the
$W$ decay products to have the $W$ mass and forces $m_t = m_{\bar t}$.
The jet permutation with the lowest $\chi^2$ is selected;
Monte Carlo studies~\cite{d0ljetm} indicate
that correct jet permutation is chosen 40\% of the time when the partons
are correctly identified in the largely untagged \D0\ sample.

\begin{figure}[hbtp]
\centerline{\protect\psfig{figure=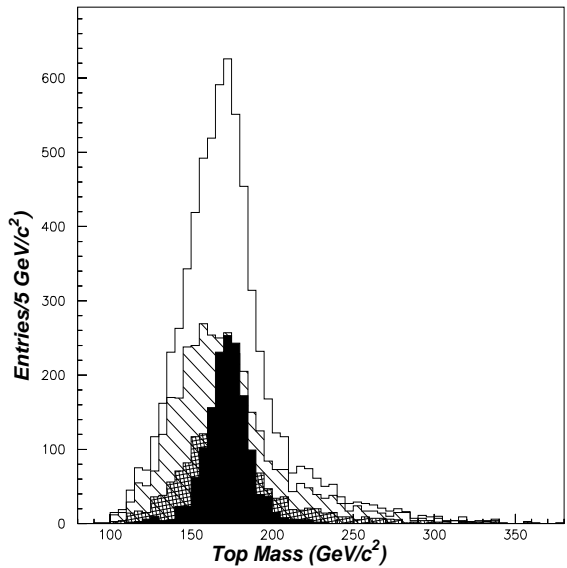,height=3in,width=3in}}
\caption{Fitted mass distribution for events with correct jet assignments (solid), events
with correctly identified partons but incorrect jet permutation (cross-hatch), events
with jet--parton mismatch (diagonal hatch), and all events (open histogram). \label{fig:rad}}
\end{figure}

The main effect of ISR, FSR, and jet combinatorics is to widen the 
reconstructed mass distribution, as shown in Fig.~\ref{fig:rad}.
Even when the wrong jet assignment is made, the reconstructed mass
distribution remains strongly correlated with the top quark mass.

\begin{figure}[hbtp]
\centerline{\protect\psfig{figure=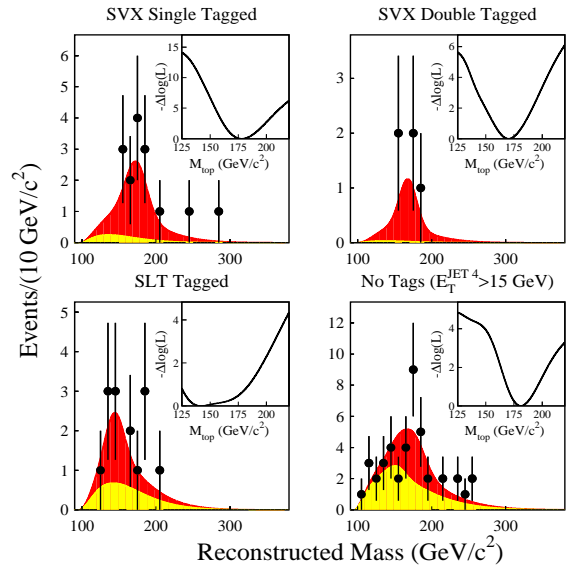,height=3in,width=3in}}
\caption{CDF Lepton~+~jets mass fits in the SVX single $b$-tag, SVX double $b$-tag,
Lepton $b$-tag (SLT), and untagged sub-samples. The light shaded area shows the expected
background contribution and the dark shaded area shows the fitted signal contribution.
The insets show the mass-dependent likelihoods for the sub-samples. \label{fig:ljmcdf}}
\end{figure}

The presence of non-top backgrounds also degrades the mass sensitivity, and
both experiments take steps to mitigate the effect of backgrounds.
The CDF lepton~+~jets mass analysis~\cite{cdfljetm} divides the event sample into four
sub-samples: SVX single $b$-tag, SVX double $b$-tag, Lepton $b$-tag, and untagged events
(see Fig.~\ref{fig:ljmcdf}).
The SVX $b$-tag samples are particularly clean, with little background expected. The SVX
double $b$-tag sample also benefits from having only two possible jet permutations,
giving rise to a noticeably narrower signal resolution. A mass-dependent likelihood is
calculated for each sub-sample and the product of the likelihoods is used to determine
the top quark mass.

\begin{figure}[hbtp]
\centerline{\protect\psfig{figure=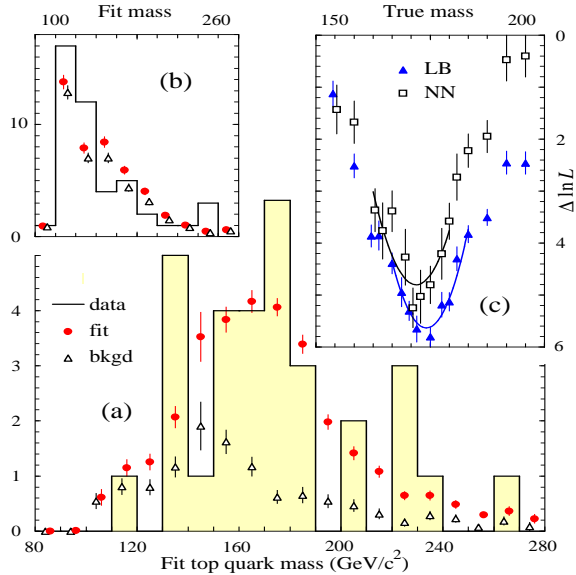,height=3in,width=3in}}
\caption{\D0\ Lepton~+~jets mass fits for (a) the signal-rich LB sample
and (b) the background-rich LB sample. The dependence of the likelihoods
on $m_t$ is shown in (c) for the LB and NN discriminants.
\label{fig:ljmd0}}
\end{figure}

The \D0\ lepton~+~jets mass analysis~\cite{d0ljetm} focuses on using kinematic variables
to separate the top signal from the background processes. By choosing variables that are
minimally correlated with the reconstructed mass, the precision of the mass
measurement is enhanced. Two different multivariate discriminants are formed from four
kinematic variables: a weighted likelihood that minimizes correlation with the reconstructed
mass (LB or ``Low Bias'') and a Neural Network output (NN). The data are binned
according to the discriminant value to create signal-rich and background-rich sub-samples.
These sub-samples are simultaneously fit to obtain the top quark mass
(see Fig.~\ref{fig:ljmd0}). Consistent results are obtained from both discriminants
and the results are combined, taking into account the expected correlation.

The CDF and \D0\ top quark mass measurements in the lepton~+~jets channel are given
in Table~\ref{tab:mass} below.

\vspace{5pt}
\par\noindent
\underline{Dilepton Channel}
\vspace{5pt}
\par\noindent
The dilepton channel requires a different approach than is used in the lepton~+~jets
channel. With two neutrinos in the final state, the event kinematics are
under-constrained (-1C) and it is not possible to solve for the top quark mass.
Nevertheless, it is possible to measure the top quark mass in the dilepton channel
with an accuracy that would be comparable to the lepton~+~jets channel if the
branching ratios were the same. 

Any quantity that is correlated with the top quark mass can, in principle,
be used to measure it.
CDF has previously measured 
$m_t=161 \pm 17\ {\rm (stat) }\pm 10\ {\rm (syst) }$ GeV/$c^2$
in the dilepton
channel by fitting the distributions of $E_b$, the $b$-quark energy,
and $m_{\ell b}$, the invariant mass between the lepton and
$b$-quark.\cite{cdfoldllm}
This result has been superceded by a new CDF dilepton mass
measurement~\cite{cdfnewllm,weiming} with a significantly
smaller error that uses the Neutrino Phase Space technique
developed by \D0.\cite{d0llm}

The CDF and \D0\ dilepton mass analyses proceed by hypothesizing a
top quark mass and determining an event weight for that hypothesis.
Two different weighting schemes have been studied:
\begin{itemize}
\item Neutrino Phase Space weighting calculates the
neutrino phase space volume consistent with the event
kinematics (CDF and \D0).\cite{d0llm}
\item Matrix Element weighting solves the event
kinematics, obtaining 0--4 physical solutions. These solutions are
then weighted by the parton distribution functions and the matrix
element for $W$ boson decay (\D0).\cite{dgk}
\end{itemize}
For both weighting schemes, the weights are summed for each jet
permutation considered (two for CDF; up to six for \D0, which
includes ISR and FSR hypotheses). An
integration over the detector resolution is performed by
repeatedly calculating the weight with the observed
event variables smeared randomly in accord with 
the detector resolutions.

\begin{figure}[hbtp]
\centerline{\protect\psfig{figure=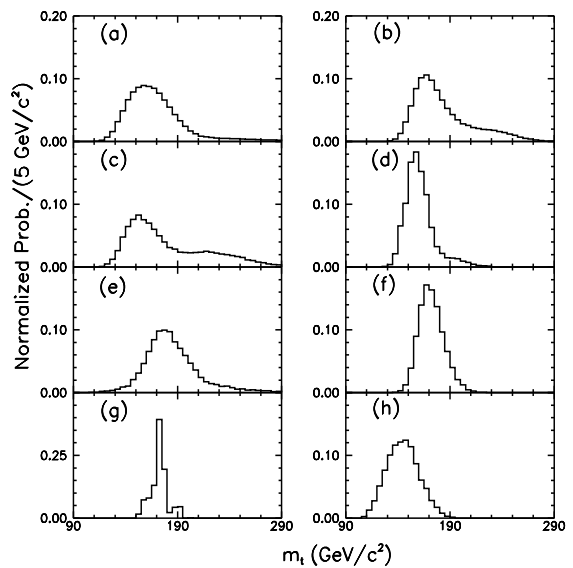,height=3in,width=3in}}
\caption{Weight functions for the CDF dilepton events using the Neutrino
Phase Space method. \label{fig:cdfweights}}
\end{figure}

\begin{figure}[hbtp]
\centerline{\protect\psfig{figure=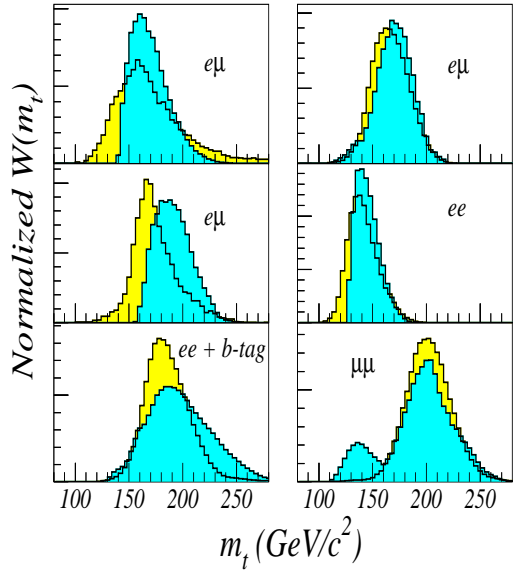,height=3in,width=3in}}
\caption{Weight functions for the \D0\ dilepton events. The lightly shaded area shows
the Neutrino Phase Space method and the darker shaded area shows the Matrix Element
method. \label{fig:d0weights}}
\end{figure}

The result of this procedure is a weight function $W(m_t)$ for each dilepton
event. Figures~\ref{fig:cdfweights}--\ref{fig:d0weights} show the weight
functions for the CDF and \D0\ dilepton events, respectively.

\begin{figure}[hbtp]
\centerline{\protect\psfig{figure=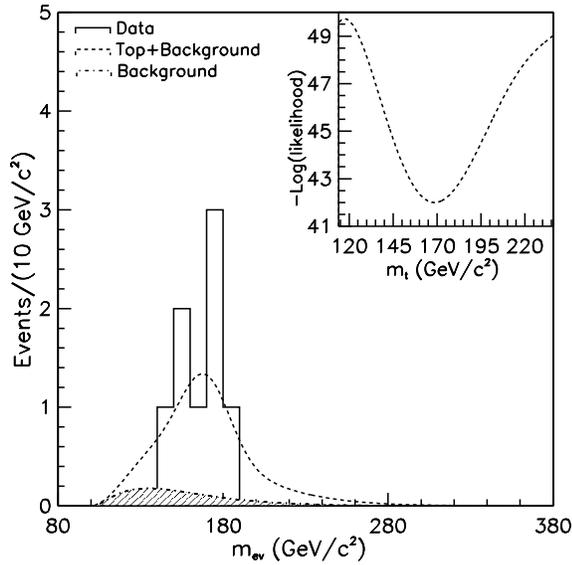,height=3in,width=3in}}
\caption{Mass distribution extracted from the CDF dilepton weight functions
and the expected distribution for the fitted top quark mass.
The hatched area shows the expected background distribution.
The inset shows the dependence of the likelihood on $m_t$. \label{fig:cdfllfit}}
\end{figure}

\begin{figure}[hbtp]
\centerline{\protect\psfig{figure=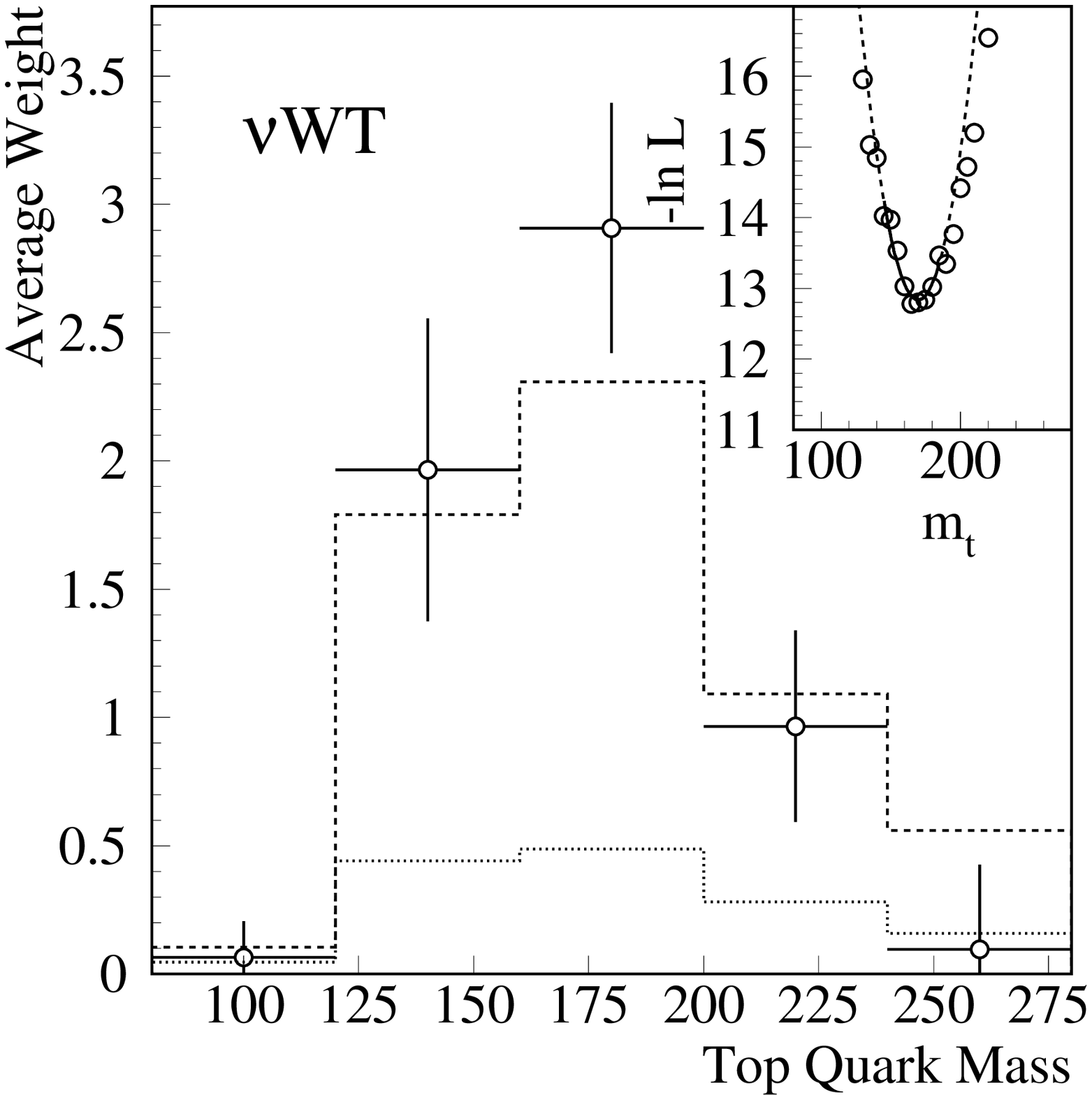,height=3in,width=3in}}
\caption{Average Neutrino Phase Space weight calculated for the 5 mass bins used
in the \D0\ mass fit (points) and the expected average weight for the fitted top quark mass.
The dotted line shows the expected averaged weight for the background. 
The inset shows the dependence of the likelihood on $m_t$.
\label{fig:d0llnu}}
\end{figure}

\begin{figure}[hbtp]
\centerline{\protect\psfig{figure=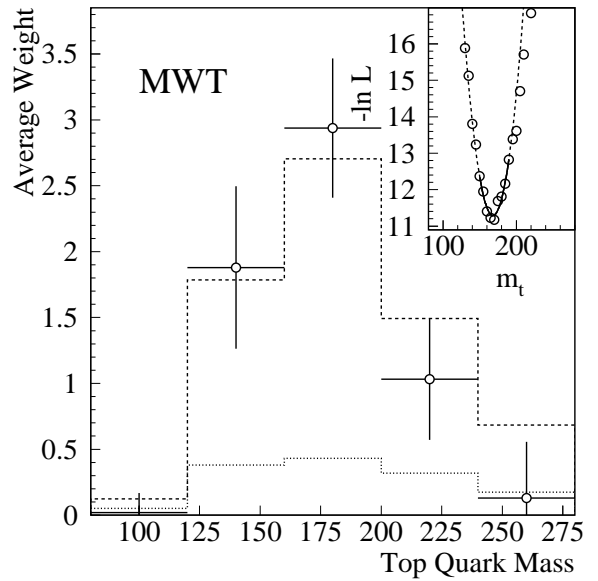,height=3in,width=3in}}
\caption{Average Matrix Element weight calculated for the 5 mass bins used in the
\D0\ mass fit (points) and the expected average weight for the fitted top quark mass.
The dotted line shows the expected average weight for the background. 
The inset shows the dependence of the likelihood on $m_t$.
\label{fig:d0llme}}
\end{figure}

The weight functions show a strong dependence on the hypothesized top quark mass.
Two different methods are used to extract the top quark mass from the weight
functions.
CDF assigns each event a mass calculated by averaging the two masses where the
weight function drops by a factor of two from its peak. Figure~\ref{fig:cdfllfit}
shows the distribution of these masses and the result of fitting this distribution.
\D0\ accounts for the shape of the weight function by integrating the weights
in five equally spaced mass bins and forming a four-dimensional vector from the
normalized weights. The top quark mass is extracted by comparing the distribution of observed
events in this four-dimensional vector space with the predicted densities for
top signal and background. The results of these fits are shown in
Figs.~\ref{fig:d0llnu}--\ref{fig:d0llme}.

The CDF and \D0\ top quark mass measurements in the dilepton channel are given
in Table~\ref{tab:mass} below.
\vspace{10pt}
\par\noindent
\underline{All-jets Channel}
\vspace{5pt}
\par
\begin{figure}[hbtp]
\centerline{\protect\psfig{figure=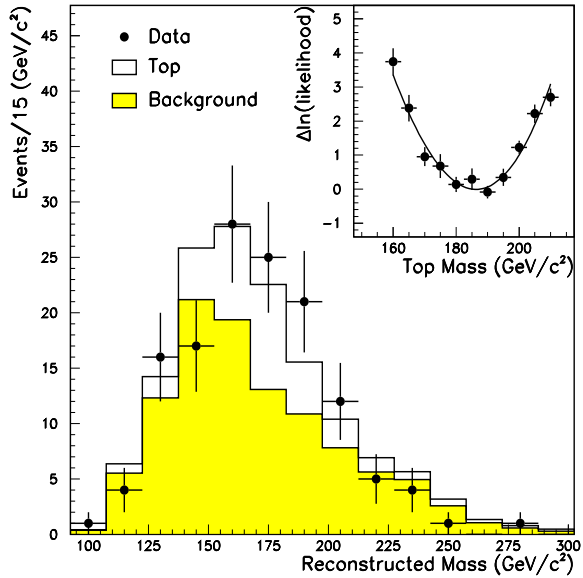,height=3in,width=3in}}
\caption{The CDF mass fit in the all-jets channel. The points show the observed mass
distribution, the solid line the fit results, and the shaded area the
expected background. The inset shows the dependence of the likelihood
on $m_t$. \label{fig:alljetm}}
\end{figure}

\noindent
The all-jets channel has a large branching fraction and, with no final state neutrinos,
can be fully measured in the detector. CDF has performed~\cite{cdfalljm}
a mass analysis in the all-jets channel that incorporates many features
of the lepton~+~jets mass analyses.
The large QCD multijet background is reduced by requiring at least one SVX $b$-tag
and imposing cuts on the kinematic variables $H_T$ and Aplanarity.
Due to the large number of jets in the final state (six), there are 30 possible
jet permutations for events with a single $b$-tag.
A three constraint (3C) kinematic fit is performed for each jet permutation and
the one with the lowest $\chi^2$ is selected.
The reconstructed mass distribution is then fit to a combination of top signal
and QCD multijet background (see Fig.~\ref{fig:alljetm}).

The CDF top quark mass measurement in the all-jets channel is given in
Table~\ref{tab:mass} below.

\begin{table}[hbt]
\begin{center}
\caption{Summary of the CDF and \D0\ top quark mass measurements.
The first error is
statistical and the second error is systematic. Units are GeV/$c^2$.}\label{tab:mass}
\vspace{0.2cm}
\begin{tabular}{|l|c|c|} 
\hline
\raisebox{0pt}[12pt][6pt]{Channel} & 
\raisebox{0pt}[12pt][6pt]{\D0} & 
\raisebox{0pt}[12pt][6pt]{CDF} \\
\hline
\raisebox{0pt}[12pt][6pt]{Lepton + Jets} &
\raisebox{0pt}[12pt][6pt]{$173.3\pm 5.6\pm 4.4$} & 
\raisebox{0pt}[12pt][6pt]{$175.9\pm 4.8\pm 5.3$} \\
\hline
\raisebox{0pt}[12pt][6pt]{Dilepton} &
\raisebox{0pt}[12pt][6pt]{$168.4\pm 12.3\pm 3.6$} & 
\raisebox{0pt}[12pt][6pt]{$167.4\pm 10.3\pm 4.8$} \\
\hline
\raisebox{0pt}[12pt][6pt]{All-jets} &
\raisebox{0pt}[12pt][6pt]{-} & 
\raisebox{0pt}[12pt][6pt]{$186.0\pm 10.0\pm 5.7$} \\
\hline
\raisebox{0pt}[12pt][6pt]{Combined} &
\raisebox{0pt}[12pt][6pt]{$172.1\pm 5.2\pm 4.9$} & 
\raisebox{0pt}[12pt][6pt]{$176.0\pm 4.0\pm 5.1$} \\ 
\hline
\raisebox{0pt}[12pt][6pt]{Tevatron} &
\multicolumn{2}{|c|}{\raisebox{0pt}[12pt][6pt]{$174.3\pm 3.2\pm 4.0$}} \\
\hline
\end{tabular}
\end{center}
\end{table}
\vspace*{3pt}

\vspace{5pt}
\par\noindent
\underline{Discussion of Top Quark Mass Results}
\vspace{5pt}
\par\noindent
Consistent values of the top quark mass are measured in all of the Standard Model
decay channels: dilepton, lepton~+~jets, and all-jets (see Table~\ref{tab:mass}).
The most precise measurement of the top quark mass is obtained by combining
these measurements, taking into account the various correlations that are present.
A joint CDF--\D0\ working group has developed a procedure for combining measurements
and produced an ``official'' combined top quark mass,
$m_t=173.8 \pm 3.2\ {\rm (stat) }\pm 3.9\ {\rm (syst) }$ GeV/$c^2$,
that was presented~\cite{ela} at the conference and used in the electroweak fits
described below.
CDF has recently revised~\cite{cdfnewllm} the systematic errors for the
lepton~+~jets and all-jets
channels; these new results are incorporated in the ``un-official''
Tevatron average shown in Table~\ref{tab:mass}. 

\begin{figure}[hbtp]
\centerline{\protect\psfig{figure=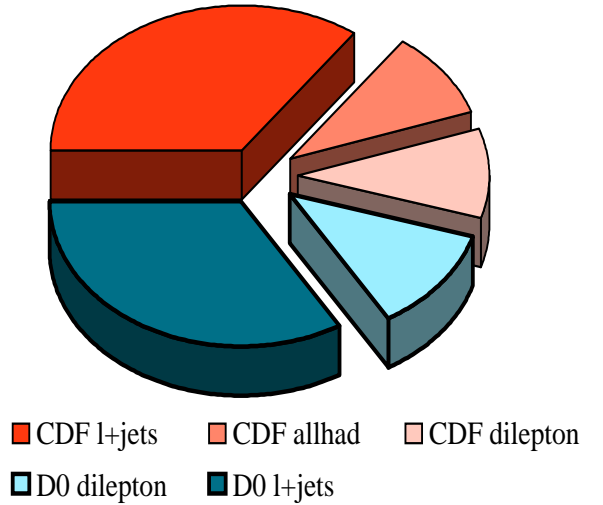,height=3in,width=3in}}
\caption{Relative weight of each channel in the top quark mass average.
The upper left pie slice is the CDF lepton~+~jets channel, followed in clockwise
order by the CDF all-jets, CDF dilepton, \D0\ dilepton, and \D0\ lepton~+~jets
channels.
\label{fig:mweights}}
\end{figure}

The good agreement among the measurements
is reflected by a $\chi^2$ probability of 75\% for the mass average.
Figure~\ref{fig:mweights} shows the relative weight of each measurement in the
``un-official'' average.
The largest weights are for the CDF and \D0\ lepton~+~jets channels; the dilepton and
all-jets channels contribute with considerably smaller weight due to their larger
statistical errors.

Combining the statistical and systematic errors in quadrature, we find
$m_t=174.3\pm 5.1$ GeV/$c^2$. Despite the low statistics in the present
data samples, the top quark is now the best measured quark mass, with
a relative precision that is under 3\%.

\begin{figure}[hbtp]
\centerline{\protect\psfig{figure=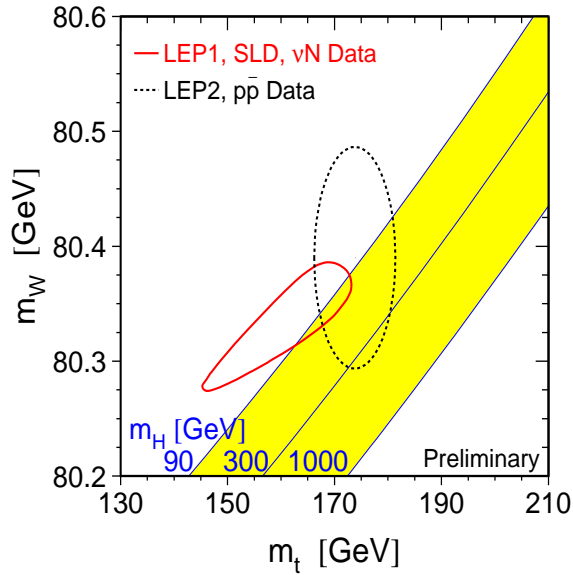,height=3in,width=3in}}
\caption{The 68\% CL contours for direct (dashed line) and indirect (solid line)
measurements of the top quark and $W$ boson masses.
The diagonal band shows the Standard Model prediction for Higgs boson masses
ranging from 90--1000 GeV/$c^2$. \label{fig:mwmt}}
\end{figure}

The precise measurement of the top quark mass constrains the electroweak radiative
corrections for the $W$ boson mass. Figure~\ref{fig:mwmt} shows the present direct
measurements of the top quark and $W$ boson masses, along with indirect constraints
from LEP, SLD,
and deep inelastic neutrino scattering. The direct and indirect measurements overlap
and have a slight preference for a light Higgs mass. 
A Standard Model fit~\cite{lewwg} of the electroweak parameters and the top quark mass
yields a Higgs mass of $m_H = 76^{+85}_{-47}$ GeV/$c^2$.

During the upcoming Tevatron Run 2, the number of produced top quarks is
expected to
increase by a factor of 30. With the increased capabilities of the upgraded
CDF and \D0\ detectors,
it should be possible to significantly improve the precision of the top
quark mass measurement. A statistical uncertainty in $m_t$ of
$\approx 0.6$ GeV/$c^2$ is obtained by scaling the present results. The dominant 
systematic errors are expected to be the uncertainties in modeling gluon radiation~\cite{orr}
and determining the jet energy scale of the detectors. The study of these uncertainties will
certainly be helped by the greatly increased data sample for Run 2, but only time
will tell if the ingenuity and hard work of those involved leads to
a comparable reduction in the systematic uncertainty on the top quark mass.
\subsection{Testing Standard Model Predictions}
Once the top quark mass is determined, the remaining properties of the top quark are
predicted by the Standard Model, opening the door for many new tests of the Standard
Model in a new regime. At present, the small statistics in the CDF and \D0\ top samples
limits the precision of these tests. The increased statistics
expected for Run 2 (and Run 3?) will allow much more precise tests of the Standard
Model predictions. The sections below describe the observation of $t\to Wb$,
the top pair production cross section, $W$ polarization in top decays, and
further tests of the Standard Model.
\vspace{5pt}
\par\noindent
\underline{Observation of $t\to Wb$}
\vspace{5pt}
\par
\begin{figure}[hbtp]
\centerline{\protect\psfig{figure=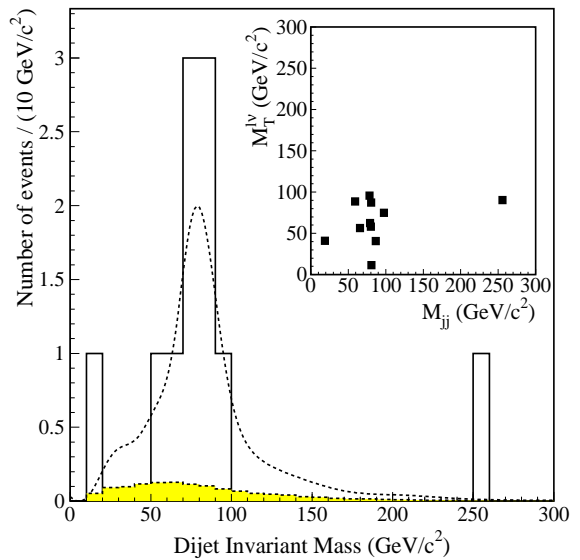,height=3in,width=3in}}
\caption{Dijet invariant mass distribution for the two highest $E_T$ untagged jets
in CDF double-tagged lepton~+~jets events. The dashed line shows the result of fitting
the mass distribution. The inset shows the correlation between the two jet mass
and the transverse mass for the event. \label{fig:mjj}}
\end{figure}

\noindent
The Standard Model predicts that top quarks will decay almost 100\% of the time
via $t\to Wb$, producing final states of $WWb\bar b$ for top pair production.
CDF has observed this final state in lepton~+~jet events that have
two $b$-quarks identified.\cite{cdfwjj}
Figure~\ref{fig:mjj} shows the invariant mass of the two highest $E_T$ jets
that are not tagged. A clear peak is observed in the mass distribution; fitting
the peak yields  
$m_{jj}=78.1 \pm 4.4\ {\rm (stat) }\pm 2.9\ {\rm (syst) }$ GeV/$c^2$, which
is in good agreement with the $W$ boson mass. The transverse mass of the $\ell\nu$
system in these events is also consistent with $W$ decay.
Thus, the final state $WWb\bar b$ has been fully reconstructed and the
existence of the Standard Model decay mode $t\to Wb$ is firmly established.

The above results are confirmed by CDF measurements~\cite{kirsten} of $R_b$, which is
expected to be close to unity,
\begin{equation}
R_b\equiv{\Gamma(t\to Wb)\over \Gamma(t\to Wq)} >0.64\hbox{\hskip .2in\rm (95\% CL)},
\label{eq:rb}
\end{equation}
and the top quark semileptonic branching ratio for a single lepton species
(either $e$ or $\mu$)
\begin{equation}
BR(t\to e/\mu X) = 0.094\pm 0.024,
\label{eq:lbr}
\end{equation}
which is expected to be 0.11 if the leptons originate from $W$ decay.
\vspace{5pt}
\par\noindent
\underline{Top Pair Production Cross Section}
\vspace{5pt}
\par\noindent
Measuring the top pair production cross section tests the Standard Model in several ways:
\begin{itemize}
\item New physics can enhance the top cross section compared to the Standard Model
predictions.
\item Top decays involving new particles can reduce
the measured top cross section that is based on Standard Model top branching ratios.
\item ``Backgrounds" from new physics processes can appear in the top sample and increase
the measured top cross section.
\end{itemize}

It is also important to measure the top cross section in as many channels as possible.
Deviations from the Standard Model predictions may be most apparent in a subset of the
top decay channels, in which case accurate measurements of the top branching ratios
provide a sensitive test.
The branching ratio measurements also have the advantage that they don't rely on the
predicted top cross section and are insensitive to some experimental uncertainties
(for example, the luminosity error).

A new measurement of the top cross section in the all-jets channel has been reported by
\D0.\cite{d0alljets,macc}
This measurement is unique in its use of jet widths to help discriminate between top signal
and QCD multijet background. Also of interest is the method of extracting the cross section
by fitting a neural network output distribution to a mixture of signal and background.

\begin{figure}[hbtp]
\centerline{\protect\psfig{figure=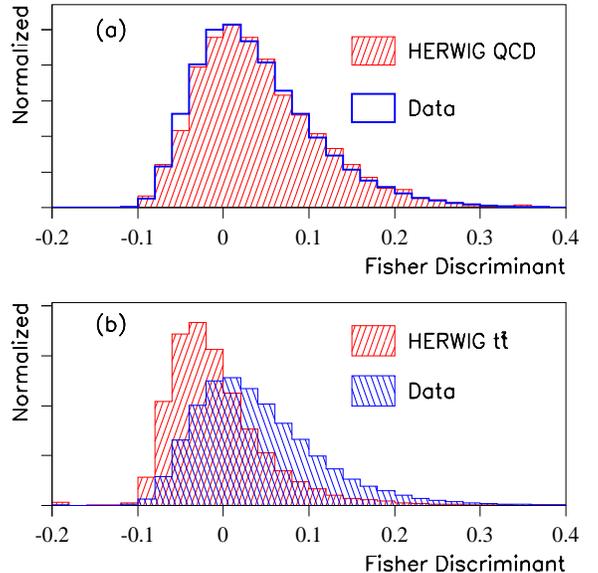,height=3in,width=3in}}
\caption{Fisher discriminant distributions showing (a) good agreement between
the \D0\ multijet data and HERWIG QCD Monte Carlo and (b) differences between \D0\
data (dominated by QCD background) and HERWIG top Monte Carlo. \label{fig:fisher}}
\end{figure}

The final state in the all-jets channel contains at least six quark
jets, while the QCD multijet background tends to have a large number of gluon jets.
Since quark jets are, on average, narrower~\cite{jetwidth} than gluon jets, the jet
widths can be used to help separate the top signal from the QCD background.
Discrimination between signal and background is achieved using a Fisher
discriminant constructed from the four narrowest jets in the event.
Figure~\ref{fig:fisher} shows that the distribution of the Fisher discriminant for
HERWIG QCD Monte Carlo events is in good agreement with the data, which is dominated
by QCD multijet production.
However, when the comparison is made between HERWIG top Monte Carlo events and the
multijet data, the top events tend to have lower values of the Fisher discriminant.

\begin{figure}[hbtp]
\centerline{\protect\psfig{figure=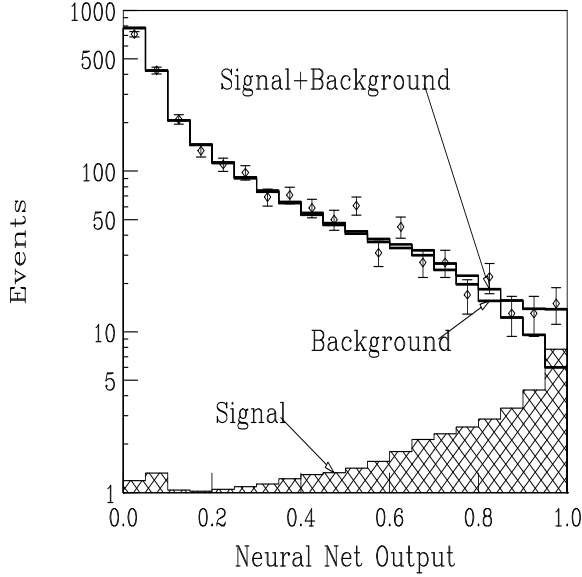,height=3in,width=3in}}
\caption{Neural network output distributions for the \D0\ data (points), top signal
(hatched), and QCD multijet background (lower histogram). Also shown is the result of
fitting the data to a combination of signal and background (upper histogram).
\label{fig:alljetsnn}}
\end{figure}

A neural network is used to combine the Fisher discriminant with 12 other
kinematic variables that also provide discrimination between the top signal and
QCD multijet background. The distribution of the neural network output for
events with a Lepton $b$-tag is shown in Fig.~\ref{fig:alljetsnn}.
The background, which is obtained from untagged events, is strongly peaked at
low values of the neural network output, while the top signal is peaked at
large values of the output. The neural network output distribution is fit to
extract the \D0\ top cross section in the all-jets channel.

\begin{figure}[hbtp]
\centerline{\protect\psfig{figure=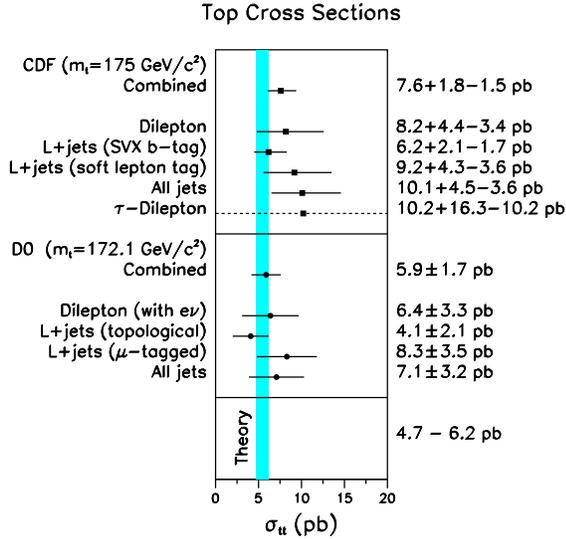,height=3in,width=3in}}
\caption{Comparison of the CDF and \D0\ cross section measurements in various top
decay channels. Also shown is the predicted top cross section. \label{fig:xsec}}
\end{figure}

Figure~\ref{fig:xsec} shows a comparison of the CDF and \D0\ top cross section measurements
in the dilepton, lepton~+~jets, all-jets, and $\tau$ dilepton
channels.\cite{cdfxsec,cdftau,d0xsec,d0alljets}
Also shown in Fig.~\ref{fig:xsec} is the range in theoretical predictions~\cite{xsecthry}
of the top cross section; these calculations are made at NLO with different approaches
to estimating the contributions from higher order diagrams.
\D0\ measures a cross section that is in good agreement with the theoretical predictions, 
while the CDF measurement is slightly (1 s.d.) higher than predicted. While the present
measurements are entirely consistent with Standard Model predictions, the rather large
experimental errors do not allow a very sensitive test at the present time. The 
increased statistics expected during Run 2 of the Tevatron collider should greatly
improve the sensitivity of the cross section measurements to new physics.
 
\vspace{5pt}
\par\noindent
\underline{$W$ Polarization in Top Decays}
\vspace{5pt}
\par\noindent
The top quark weak interaction coupling can be probed by measuring the $W$ polarization
in top decays. The Standard Model predicts that both left-handed and longitudinal
$W$ bosons are produced in top decay, with the fraction $F_0$ that are longitudinal
given by
\begin{equation}
\begin{array}{rcl}
F_0 &=& (1 + 2m_W^2/m_t^2)^{-1} \\
&\approx& 0.70.
\end{array}
\label{eq:wpol}
\end{equation}

CDF has performed a preliminary analysis~\cite{kirsten} of the $W$ polarization
based on fitting the distribution of transverse momenta ($p_T$) of leptons in the
lepton~+~jets and dilepton channels. The $p_T$ distribution is expected to be
softer for left-handed $W$ bosons since the lepton decay angular
distribution is peaked in the direction opposite the boost.

\begin{figure}[hbtp]
\centerline{\protect\psfig{figure=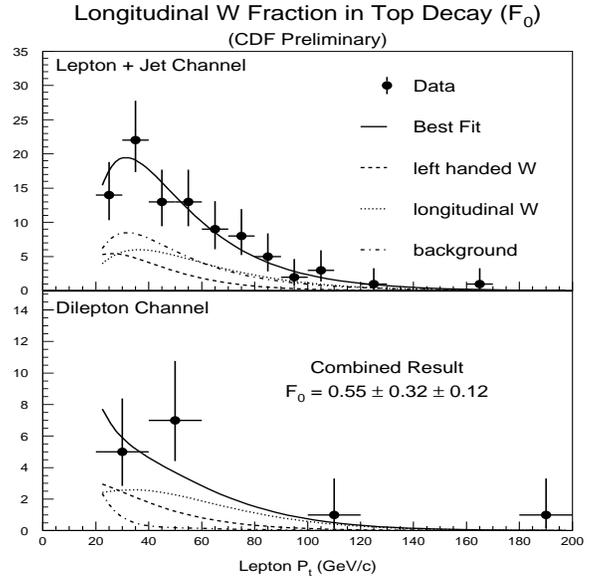,height=3in,width=3in}}
\caption{Distribution of the lepton $p_t$ in CDF lepton~+~jets and dilepton
event samples. Also shown is the result of fitting this distribution to a mixture
of left-handed and longitudinal $W$ polarizations. \label{fig:wpol}}
\end{figure}

Figure~\ref{fig:wpol} shows the result of fitting the lepton $p_T$ distributions.
Given that $F_0$ is constrained by definition to lie in the range 0--1, the present
CDF result $F_0=0.55\pm 0.32\ {\rm (stat) }\pm 0.12\ {\rm (syst) }$ is not sufficiently
precise to test the Standard Model prediction. What is noteworthy
about this result is that the systematic error is sufficiently small that it should be
possible to make an interesting measurement in Run 2. Furthermore, there exist other
methods for determining the $W$ polarization, such as measuring the decay angle of
the lepton in the $W$ rest frame, that may prove useful in making a precise measurement
of $F_0$.

\vspace{15pt}
\par\noindent
\underline{Further Tests of the Standard Model}
\vspace{5pt}
\par\noindent
Many other tests of the Standard Model are possible in addition to those described
above. They include:
\begin{itemize}
\item Measurement of the single top production cross section, which can be used to extract
$|V_{tb}|$ and test the expectation that the CKM matrix is a
$3\times 3$ unitary matrix.
\item Measurement of $t\bar t$ spin correlations, which tests the prediction that top
will decay before hadronizing.
\item Measurement of kinematic distributions that are sensitive to new physics. For
example, a topcolor~\cite{topcolor} $Z^\prime$ has a large branching ratio to $t\bar t$
and would appear as a peak in the $m_{t\bar t}$ distribution. CDF~\cite{cdfkin} and
\D0~\cite{d0ljetm} have studied the distributions of several kinematic variables
for the top candidates in the lepton~+~jets channel and find no significant
deviations from the Standard Model predictions with the present statistics.
\item Search for new particles in top quark decays. For example, a charged Higgs
that is lighter than the top quark could be seen in $t\to H^+ b$ decays. Another
possibility is the production of the stop squark in $t\to \tilde t \tilde \chi$
decays.
CDF has
published~\cite{cdfhiggs} the results of a direct search for the charged Higgs in the
$H^+\to \tau\nu$ decay channel. More stringent limits will come from indirect
searches that exclude regions of the parameter space where the $t\to H^+ b$
branching ratio is larger than can be accommodated given the measured top cross
sections. Earlier indirect searches by CDF and \D0\ did not take into
account the decay $H^+\to t^* \bar b\to Wb\bar b$ and are being revised; no new results
were presented at the conference.
\item Search for top decays into final states other than $t\to Wb$. For example,
CDF has set limits on two FCNC decay modes: $BR(t\to Zq) < 0.33$ and
$BR(t\to \gamma q) < 0.032$ (95\% CL).\cite{cdfrare}
\end{itemize}
All of the above tests will benefit from the increased statistics expected in the
next run.
\section{Measurement of the \bsg\ Branching Ratio}
The measurement of the branching ratio for \bsg\ is the first of several
topics related to the production and decay of the ``lighter'' heavy quarks, bottom and
charm.

\begin{figure}[hbtp]
\centerline{\protect\psfig{figure=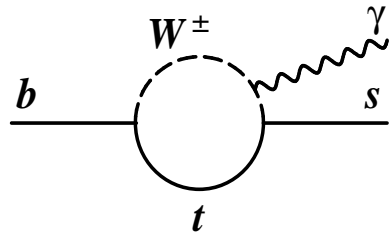,height=1.2in,width=2in}}
\caption{One of the loop diagrams that contributes to the FCNC decay
\bsg.
\label{fig:penguin}}
\end{figure}

Loop diagrams, such as the one shown in
Fig.~\ref{fig:penguin}, can give rise to FCNC decays of the $B$ meson.
Since the particles appearing in the loop are not limited to those already
discovered, such decays are sensitive to new physics beyond the Standard Model.
For example, the $W$ boson in Fig.~\ref{fig:penguin} could be replaced by
a charged Higgs, possibly leading to a deviation between the measured
\bsg\ branching ratio and the Standard Model prediction. 
While such a deviation would be evidence for new physics beyond the Standard
Model, the converse is not necessarily true. In SUSY models, for example,
there can be cancelations between the charged Higgs loop and loops
involving supersymmetric particles.
Thus, limits on new physics inferred from the measured \bsg\
branching ratio are necessarily model dependant.

The signature for \bsg\ decays is a photon with $E_\gamma\approx m_b/2$
in the $b$-hadron rest frame. 
The photon energy spectrum is broadened by the Fermi motion of the $b$-quark within the
$b$-hadron and by the boost of the $b$-hadron with respect to the lab frame.

The \bsg\ branching ratio has previously been measured by CLEO~\cite{cleobsg1}
(see Table~\ref{tab:bsg} below).
Two new measurements of this branching ratio were presented at the conference: an
updated CLEO measurement~\cite{cleobsg2} with a 60\% larger data sample and a lower
photon energy threshold, and a new ALEPH measurement~\cite{alephbsg,litke}
using $Z\to b\bar b$ decays.
A new calculation of the Standard Model branching ratio was also presented.\cite{neubert}

The updated CLEO analysis takes two approaches to suppressing the continuum background.
The first approach is to feed eight event-shape variables into a neural network and to
weight events by the expected signal fraction for the value of the neural network output.
The second approach is to attempt reconstruction of the $B$ decay and is used for events
with a plausible reconstructed $B\to X_s\gamma$ final state.
The final states considered have either a $K^\pm$ or a $K_S^0\to\pi^+\pi^-$ decay
and 1--4 $\pi$'s with at most 1 $\pi^0$. A second neural network combines the output of
the event-shape neural network with two measures of the consistency
of the reconstructed final state with $e^+e^-\to B\bar B$.
The output of this second neural network is then used to calculate event weights as before.

\begin{figure}[hbtp]
\centerline{\protect\psfig{figure=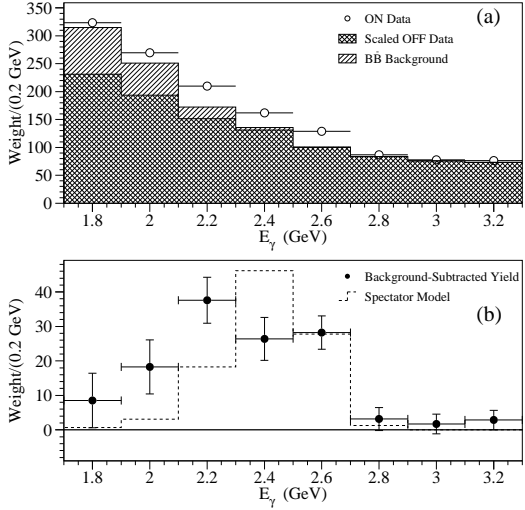,height=3in,width=3in}}
\caption{The CLEO photon energy spectrum for (a) selected events (points), continuum
background (cross-hatched), and $B\bar B$ background (diagonal hatched); and (b)
background subtracted data (points) and a Monte Carlo prediction for the shape
of the \bsg\ signal (dashed line). \label{fig:cleobsg}}
\end{figure}

Figure~\ref{fig:cleobsg} shows the weighted photon energy spectrum. The \bsg\ signal is
seen as an excess of event weights over the continuum and $B$ backgrounds.
CLEO uses the number of events with photon energies
in the 2.1--2.7 GeV range to determine the \bsg\ branching ratio
(see Table~\ref{tab:bsg} below). The previous CLEO analysis used a more
restricted photon energy range (2.2--2.7 GeV); the increased range in photon energy is largely
responsible for the larger branching ratio in the updated result.

\begin{figure}[hbtp]
\centerline{\protect\psfig{figure=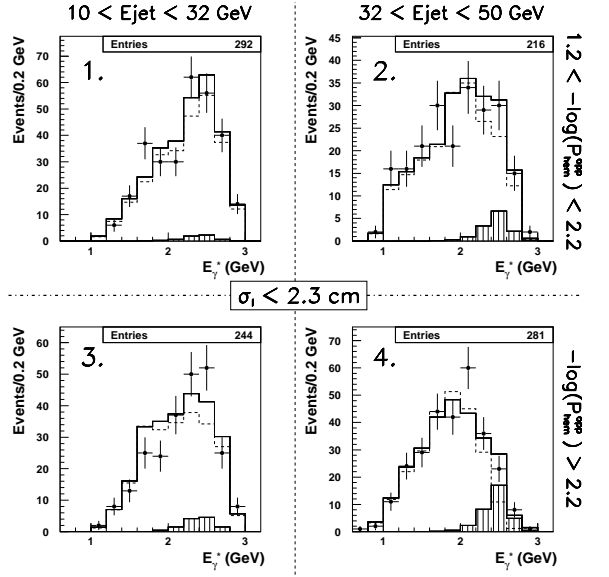,height=3in,width=3in}}
\caption{The distributions of $E_\gamma^*$ (points)
for the ALEPH $\gamma$-like sub-samples with (1) low jet energy and low $b$-purity,
(2) high jet energy and low $b$-purity, (3) low jet energy and high $b$-purity,
and (4) high jet energy and high $b$-purity. Also shown are the results of the
fit (solid line), the fitted signal (shaded area), and the expected background
before the fit (dashed line). \label{fig:aleph1}}
\end{figure}

ALEPH selects $Z$ boson decays that have a high energy photon candidate and event kinematics
that are consistent with \bsg\ decay of a $b$-hadron ($B$, $B_s$, or
$\Lambda_b$). An inclusive reconstruction algorithm is used to identify the decay
products of the $b$-hadron and calculate $E_\gamma^*$, the photon energy in the
$b$-hadron rest frame.
ALEPH then divides its data into eight sub-samples based on three criteria:
the electromagnetic cluster is ``$\gamma$-like'' or ``$\pi^0$-like'', the jet
containing the photon has high or low jet energy, and the hemisphere opposite
the $b$-hadron has high or low ``$b$-purity.''
The signal and background normalizations are determined by a simultaneous
fit to the $E_\gamma^*$ distributions for the eight sub-samples.
The fits for the $\gamma$-like sub-samples are shown in Fig.~\ref{fig:aleph1}.
The \bsg\ signal is most clearly seen in sub-sample 4, which has the
highest efficiency for \bsg\ decays (see Fig.~\ref{fig:aleph2}).
The ALEPH measurement of the \bsg\ branching ratio is given
in Table~\ref{tab:bsg} below.

\begin{figure}[hbtp]
\centerline{\protect\psfig{figure=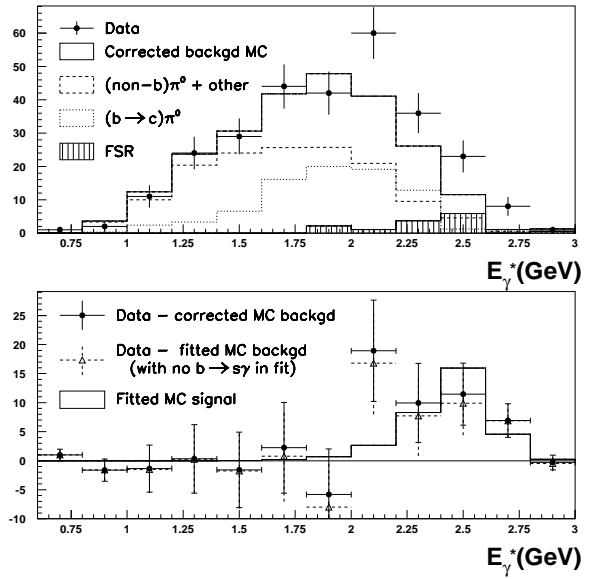,height=3in,width=3in}}
\caption{The distribution of $E_\gamma^*$ for the ALEPH $\gamma$-like sub-sample
with high jet energy and high $b$-purity (sub-sample 4 in Fig.~\ref{fig:aleph1}).
The top figure shows the data and
the various background contributions. The bottom figure shows the background
subtracted data and the fitted \bsg\ signal.
Also shown are the background-subtracted data points where the background
normalization was determined by fitting the data with no
\bsg\ included in the fit. \label{fig:aleph2}}
\end{figure}

A new theoretical calculation~\cite{neubert} of the Standard Model branching ratio
for \bsg\ was presented that is in good agreement with the CLEO and ALEPH measurements.
A precise measurement of the photon energy spectrum would help reduce the theoretical
uncertainty in the \bsg\ branching ratio measurements and also provides a way to probe
the momentum distribution of $b$-quarks inside $B$ mesons. 
To illustrate the importance of determining the photon energy spectrum, note that both CLEO
and ALEPH see small ``statistical fluctuations'' in the photon energy bin at
$\approx 2$ GeV; if confirmed with higher statistics this
would indicate a broader photon energy distribution than is presently being modeled and
increase the measured \bsg\ branching ratios.

\begin{table}[hbt]
\begin{center}
\caption{Summary of the CLEO and ALEPH measurements of the branching ratio for
\bsg. The Standard Model prediction is also given.
The first error is statistical, the second error is systematic,
and the third error is the model dependence (new CLEO only).}\label{tab:bsg}
\vspace{0.2cm}
\begin{tabular}{|l|l|} 
\hline
\raisebox{0pt}[12pt][6pt]{Experiment} & 
\raisebox{0pt}[12pt][6pt]{BR(\bsg)} \\
\hline
\raisebox{0pt}[12pt][6pt]{CLEO (old)} &
\raisebox{0pt}[12pt][6pt]{$(2.32\pm 0.57\pm 0.35)\times 10^{-4}$} \\
\hline
\raisebox{0pt}[12pt][6pt]{CLEO (new)} &
\raisebox{0pt}[12pt][6pt]{$(3.15\pm 0.35 \pm 0.32 \pm 0.26)\times 10^{-4}$} \\
\hline
\raisebox{0pt}[12pt][6pt]{ALEPH} &
\raisebox{0pt}[12pt][6pt]{$(3.11\pm 0.80\pm 0.72)\times 10^{-4}$} \\ 
\hline
\raisebox{0pt}[12pt][6pt]{Standard Model} &
\raisebox{0pt}[12pt][6pt]{$(3.29\pm 0.33)\times 10^{-4}$} \\
\hline
\end{tabular}
\end{center}
\end{table}
\vspace*{3pt}
\section{$b$-Quark Production in $p\bar p$ and $ep$ Collisions}
\begin{figure}[hbtp]
\centerline{\protect\psfig{figure=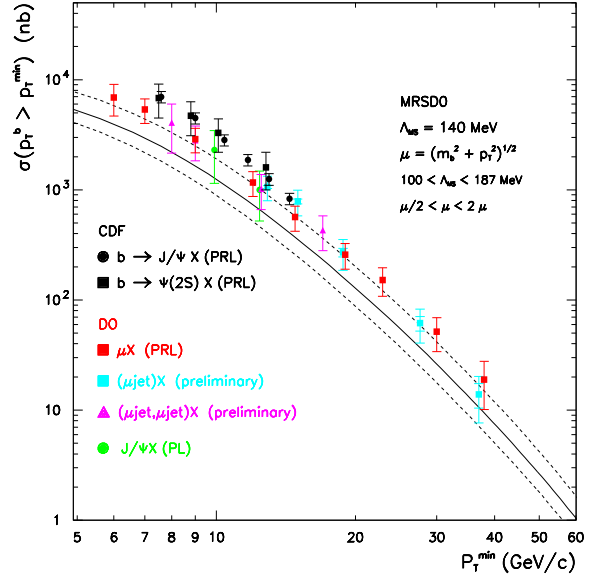,height=3in,width=3in}}
\caption{The inclusive central ($|y|<1$) $b$-quark production cross section measured
by CDF and \D0.
The curves show the NLO QCD prediction (solid line) and the effect of varying
$\mu$ between $\mu_0/2$ and $2\mu_0$ and $m_b$ between 4.5 and 5.0 GeV/$c^2$
(dashed lines).  \label{fig:bcent}}
\end{figure}

The inclusive central $b$-quark production cross section has been measured in
$p\bar p$ collisions by CDF~\cite{cdfbcent}
and \D0~\cite{d0bcent} (see Fig.~\ref{fig:bcent}).
The measured cross section has the same $p_T$ dependence
as the NLO QCD calculation~\cite{mnr} 
but is a factor of 2--3 higher than expected for
$\mu = \mu_0 = \sqrt{m_b^2+p_t^2}$ and $m_b = 4.75$ GeV/$c^2$.
A variety of new results on $b$-quark production were presented 
at the conference.

\begin{figure}[hbtp]
\centerline{\protect\psfig{figure=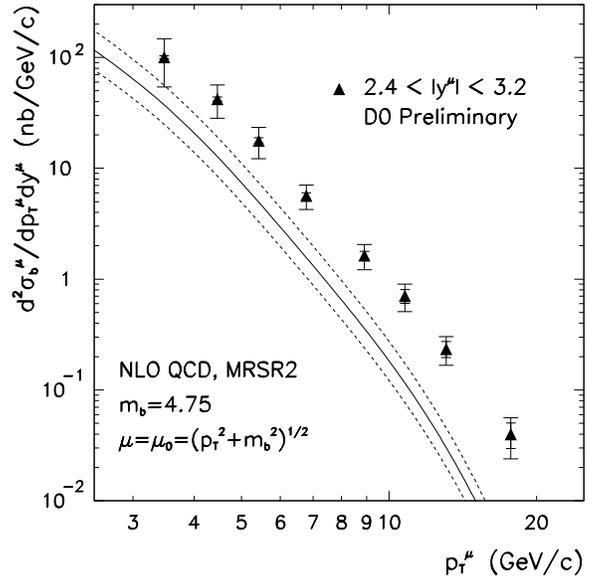,height=3in,width=3in}}
\caption{The $p_T$ spectrum of forward muons ($2.4 < |y^\mu|<3.2$) from $b$-quark
decay measured by \D0.
The curves show the NLO QCD prediction (solid line) and the effect of varying
$\mu$ between $\mu_0/2$ and $2\mu_0$ and $m_b$ between 4.5 and 5.0 GeV/$c^2$ (dashed lines). \label{fig:bforward}}
\end{figure}

\D0\ has observed $b$-quark production in the forward region by identifying muons in its
small angle muon spectrometer.\cite{stichelbaut}
The expected contribution from in-flight $\pi/K$
decay is subtracted and a NLO QCD Monte Carlo~\cite{mnr} is used to determine the fraction of
detected muons that originate from $b$-quark decay (as opposed to $c$-quark decay).
Figure~\ref{fig:bforward} shows the $p_T^\mu$ spectrum for muons in the rapidity interval
$2.4<|y^\mu|<3.2$. 
The NLO QCD prediction is consistently smaller than the data and falls somewhat more
steeply with $p_T^\mu$ than the data.
 
\begin{figure}[hbtp]
\centerline{\protect\psfig{figure=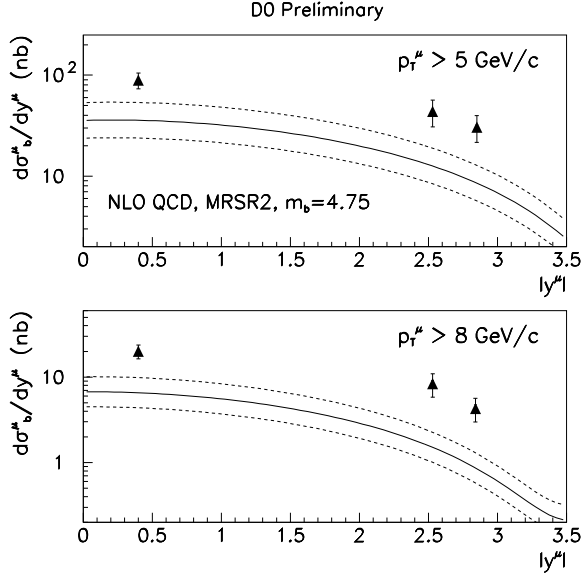,height=3in,width=3in}}
\caption{Rapidity dependence of muons from $b$-quark decay measured by \D0\ for
$p_T^\mu>5$ GeV/$c$ (top figure) and $p_T^\mu>8$ GeV/$c$ (bottom figure).
The curves show the NLO QCD prediction (solid line) and the effect of varying
$\mu$ between $\mu_0/2$ and $2\mu_0$ and $m_b$ between 4.5 and 5.0 GeV/$c^2$
(dashed lines). \label{fig:brap}}
\end{figure}

The \D0\ forward measurements are combined with the \D0\ central measurements ($|y^\mu|<0.8$)
to show the rapidity dependence of the cross section for muons from $b$-quark decay
for two different $p_T^\mu$ ranges (see Fig.~\ref{fig:brap}).
For $p_T^\mu>5$ GeV/$c$, the data/theory ratio is $2.5\pm 0.5$ in the central region
and increases to $3.6\pm 0.6$ in the forward region. For $p_T^\mu>8$ GeV/$c$,
the data/theory ratio is $3.1\pm 0.6$ in the central region and increases to
$5.1\pm 1.1$ in the forward region. (The uncertainties in the above ratios only
reflect the experimental errors.)

\begin{figure}[hbtp]
\centerline{\protect\psfig{figure=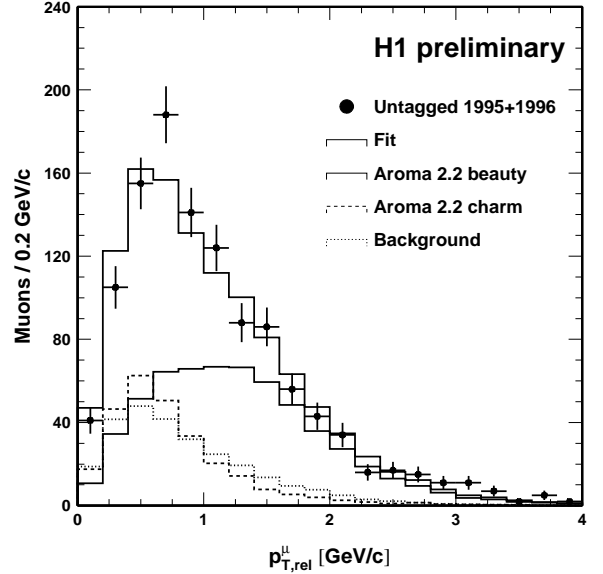,height=3in,width=3in}}
\caption{The distribution of $p_T^{rel}$, the transverse momentum of the muon
with respect to the jet axis, for H1 data (points). Also shown is the result
of fitting the data to a mixture of background, $c$-quark production, and
$b$-quark production. \label{fig:herab}}
\end{figure}

The first observation of $b$-quark production in $ep$ collisions at HERA was
reported~\cite{yorgos} at the conference. The production of $b$-quarks at
HERA is predominately through the photon--gluon fusion process, $\gamma g \to b\bar b$.
H1 selects $b$-quark candidates by requiring a muon be detected within a jet.
The $b$-quark production cross section is determined by fitting the $p_T^{rel}$
distribution, where $p_T^{rel}$ is the transverse momentum of the muon with
respect to the jet axis (see Fig.~\ref{fig:herab}).
For the kinematic region $Q^2<1$ GeV$^2$, $0.1 < y < 0.8$, $p_T^\mu >2.0$ GeV/$c$,
and $35^\circ < \theta^\mu < 130^\circ$, H1 measures
\begin{equation}
\sigma(ep\to b\bar bX)= 0.93\pm 0.08\ {\rm (stat) }\ ^{+0.17}_{-0.07}\ {\rm (syst)\ nb.}
\label{eq:h1b}
\end{equation}
A leading order Monte Carlo (AROMA 2.2) predicts this cross section to be 0.19 nb,
approximately a factor of five lower than is observed.

\begin{figure}[hbtp]
\centerline{\protect\psfig{figure=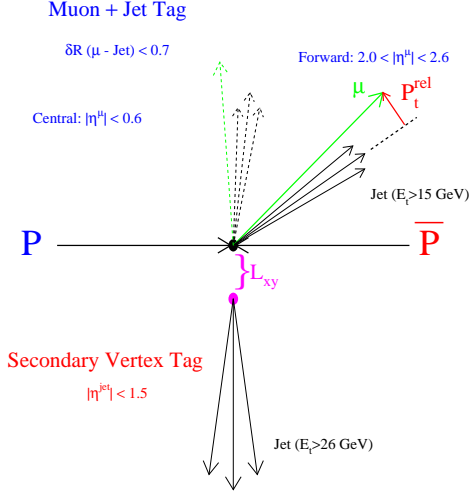,height=3in,width=3in}}
\caption{Diagram showing the CDF technique used to measure rapidity correlations.
\label{fig:bbcor}}
\end{figure}

Another way to test NLO QCD predictions for $b$-quark production is to study $b\bar b$
correlations. Previous CDF~\cite{cdfbbcor} and \D0~\cite{d0bbcor} studies of $b\bar b$
correlations examined the correlations in azimuthal angle. New CDF results on 
$b\bar b$ rapidity correlations were presented at the conference.\cite{stichelbaut}
The measurement of $b\bar b$ rapidity correlations where one $b$ is central and the
other $b$ is forward probes the gluon pdf at large $x$ ($x_{high}\approx 0.25$, 
$x_{low}\approx 0.025$).
CDF requires that one $b$-quark $(b_1)$ be identified by having a muon near a jet
consistent with semileptonic $b$-quark decay and a second $b$-quark $(b_2)$ be
identified by having a SVX $b$-tag. The signal is extracted
by fitting the muon $p_T^{rel}$ and the SVX pseudo-$c\tau$ distributions for
the cases where $b_1$ lies in either the central $(|y_{b_1}|<0.6)$ or forward
$(2.0<|y_{b_1}|<2.6)$ regions (see Fig.~\ref{fig:bbcor}).

\begin{figure}[hbtp]
\centerline{\protect\psfig{figure=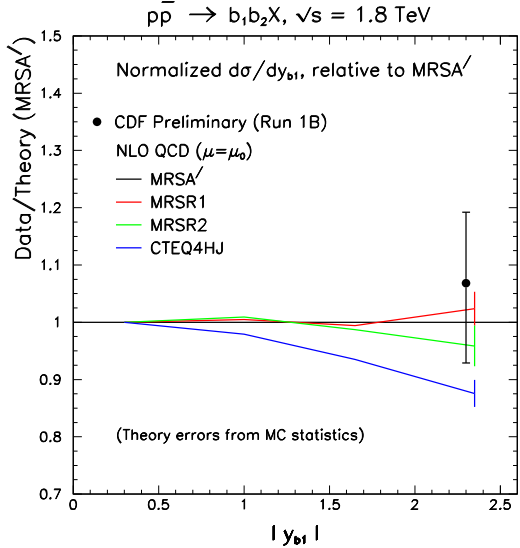,height=3in,width=3in}}
\caption{The ratio between the CDF measured rapidity correlation and the NLO QCD prediction
using the MRSA$^\prime$ pdf. Also shown is the effect of various pdf choices on this
ratio. \label{fig:bbpdf}}
\end{figure}

The rapidity correlation is measured by taking the ratio of the cross sections for
$b_1$ in the forward and central regions:
\begin{equation}
R = {\sigma(p\bar p\to b_1b_2X;\ 2.0 <|y_{b_1}| < 2.6) \over
\sigma(p\bar p\to b_1b_2X;\ |y_{b_1}| < 0.6)}. 
\label{eq:bbcor1}
\end{equation}
For the kinematic region $p_T(b)>25$ GeV/$c$, $|y_{b_2}|<1.5$, and
$\delta\phi(b\bar b)>60^\circ$, CDF finds:
\begin{equation}
R_{data} = 0.361\pm 0.41\ {\rm (stat) }\ ^{+0.011}_{-0.023}\ {\rm (syst)},
\label{eq:bbcor2}
\end{equation}
in good agreement with the NLO QCD prediction using the MRSA$^\prime$ pdf:
\begin{equation}
R_{theory} = 0.338^{+0.014}_{-0.097}.
\label{eq:bbcor3}
\end{equation}
At the present time, the rapidity correlation is not measured with sufficient
precision to distinguish between various pdf choices due to the large statistical
error (see Fig.~\ref{fig:bbpdf}).

To summarize the $b$-production results, we note that there continue to be
discrepancies between the measured $b$-quark production cross section and the
NLO QCD prediction in $p\bar p$ collisions at the Tevatron. A new
measurement of forward $b$ production shows that the discrepancy is larger
in the forward region and the measured $p_T$ spectrum falls more slowly
than the NLO QCD prediction in this region.
Proposed explanations for these discrepancies include a new theoretical
calculation based on a variable flavor number scheme~\cite{vflavor}
and a harder $b$-quark fragmentation function.\cite{bfrag}
The first measurement of the $b$-quark production cross section by H1
at HERA indicates that a similar discrepancy may be present in $ep$
collisions.
One area of agreement between data and NLO QCD is in the
$b\bar b$ rapidity correlations for $b$-quark jets.
\section{$b$-Quark Fragmentation Fractions}
The $b$-quark fragmentation fractions are used to describe the non-perturbative
fragmentation of a $b$-quark to a weakly decaying $b$-hadron.
They are an important ingredient in constructing accurate models of $b$-quark 
production in high energy collisions and must be determined experimentally.

The fragmentation fractions $f_{B^0}$, $f_{B^+}$, $f_{B_s}$, and $f_{\Lambda_b}$ are
defined as the fraction of $b$-quarks that fragment to $B^0$, $B^\pm$, $B_s$, and
$b$-baryon, respectively.
The fraction of $b$-quarks that fragment to $B_c$ is expected to be small.\cite{bcfrag}
New CDF and DELPHI measurements of the $b$-quark fragmentation fractions 
were presented~\cite{ugo} at the conference and are described below.

\begin{figure}[hbtp]
\centerline{\protect\psfig{figure=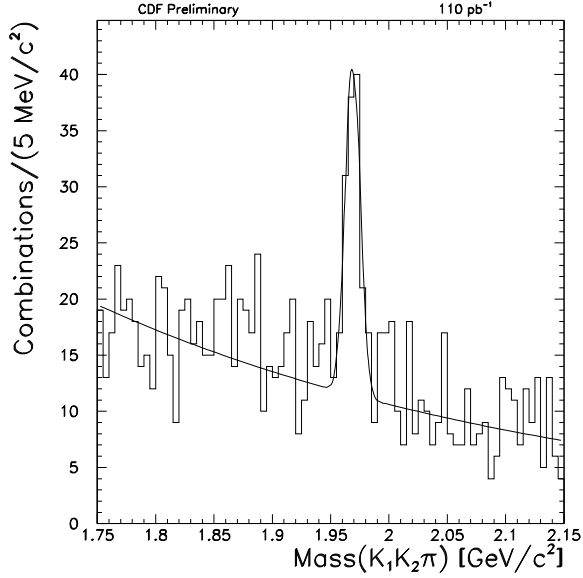,height=3in,width=3in}}
\caption{The CDF $KK\pi$ mass distribution for events consistent
with the decay hypothesis $B_s\to e^\pm\nu D_s^\mp X$, $D_s\to \phi\pi\to KK\pi$.
The curve shows a fit to the $D_s$ peak.
\label{fig:bscdf}}
\end{figure}

CDF has reconstructed the inclusive semileptonic decay channels associated with
the weakly decaying $b$-hadrons:
\begin{equation}
\begin{array}{rcl}
B^+ &\to& e^+\nu \bar{D^0} X\\
B^0 &\to& e^+\nu D^{(*)-} X\\
B_s^0 &\to& e^+\nu D_s^- X\\
\Lambda_b &\to& e^+\nu \Lambda_c^- X.
\end{array}
\label{eq:CDFBdecays}
\end{equation}
The relative rates for these decay channels can be used to determine ratios of
the fragmentation fractions.
To illustrate the technique, consider the decay $B_s^0 \to e^+\nu D_s^-$.
Reconstructing a $D_s$ uniquely identifies the weakly decaying $b$-hadron
to be a $B_s$.
Figure~\ref{fig:bscdf} shows the $KK\pi$ mass distribution for events
consistent with the decay hypothesis $B_s\to e^\pm\nu D_s^\mp X$, 
$D_s\to \phi\pi\to KK\pi$.
A clear $D_s$ peak is seen in this channel; the rate is then determined by
fitting the peak with the expected line shape.

The extraction of $b$-quark fragmentation fractions is complicated by cross contamination
among the channels. For example, $B^+\to e^+\nu\bar D^{**0}X$ with $\bar D^{**0}\to D^-\pi^+$
contaminates the $B^0\to e^+\nu D^-X$ channel. By accounting for cross contamination and
applying the constraint $f_{B^0} + f_{B^+} + f_{B_s} + f_{\Lambda_b} = 1$, the $b$-quark
fragmentation fractions are obtained. Subsequent to the conference, CDF completed revisions
to this analysis and the new results are given in Table~\ref{tab:bfrag}.\cite{wendy}
The ratio $f_{B^0}/f_{B^+} = 0.770\pm0.19$ is consistent with the expectation
$f_{B^0}=f_{B^+}$.

DELPHI has used inclusive properties of $b$-quark jets to determine the fragmentation
fractions. Two different techniques are used: correlations in the particles
produced in association with the $b$-hadron and the charge of the $b$-hadron decay products.

To conserve flavor, a $B_s$ ($\bar b s$) meson must be produced in association with an
$\bar s$-quark, which is likely to result in a $K^+$ ($\bar s u$) meson that is correlated with
the $b$-quark direction. By contrast, a $B^+$ ($\bar b u$) hadron must be produced in
association with a $\bar u$ quark, which is likely to result in either a non-strange
particle or a $K^-$ ($\bar u s$) meson that is correlated with the $b$-quark direction.

\begin{figure}[hbtp]
\centerline{\protect\psfig{figure=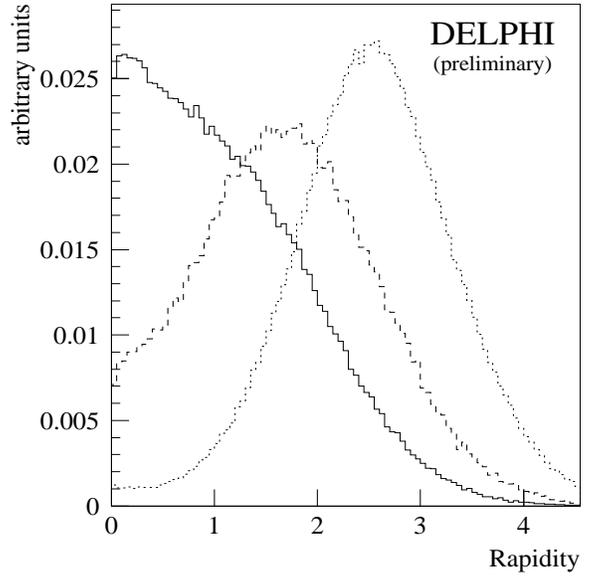,height=3in,width=3in}}
\caption{Distribution of rapidity with respect to the thrust axis for particles from $b$-hadron
decay (dotted), the fragmentation ``partner'' of the $B$-hadron (dashed), and other
fragmentation tracks (solid). It is the fragmentation partner that is expected to exhibit
a flavor correlation with the $b$-hadron. All distributions are from simulations. \label{fig:rapcor}}
\end{figure}

\begin{figure}[hbtp]
\centerline{\protect\psfig{figure=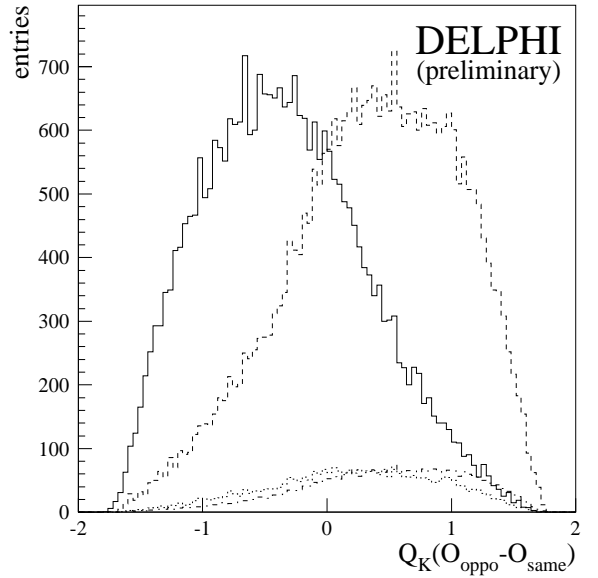,height=3in,width=3in}}
\caption{Distributions of the DELPHI flavor correlation variable $Q_K(O_{oppo}-O_{same})$ for
$B_s$ mesons (solid) and non-strange $b$-hadrons (dashed), where $Q_K$ is the charge of
the $K$ meson and $O_{oppo}$ ($O_{same}$) is the output of a neural network trained to
give +1 for $\bar b$-quarks and -1 for $b$-quarks in the opposite (same) hemisphere as
the $K$ meson. The distribution of the flavor correlation variable for non-strange
$b$-hadrons is shown scaled down in size for neutral (dotted) and charged (dash-dotted)
$b$-hadrons. All distributions are from simulations. \label{fig:qkcor}}
\end{figure}

The fraction $f_s^\prime$ of $b$-quarks that fragment to a $B_s$ or one of its excitations
is determined by fitting the joint distribution of two variables that identify flavor correlations.
The first variable is the rapidity with respect to the thrust axis; it is sensitive to where the
$K$ meson was produced in the fragmentation chain (see Fig.~\ref{fig:rapcor}).
The second variable is the product of the $K$ charge and a variable that is sensitive to
whether a $b$-quark or $\bar b$-quark was produced; it is a measure of the flavor correlation
described above (see Fig.~\ref{fig:qkcor}).
The fragmentation fraction $f_{B_s}$ is determined by correcting $f_s^\prime$ for
$B_s^{**}\to B^{(*)}_{u,d} K$ decays that do not give rise to a weakly decaying $B_s$ meson.
 
\begin{figure}[hbtp]
\centerline{\protect\psfig{figure=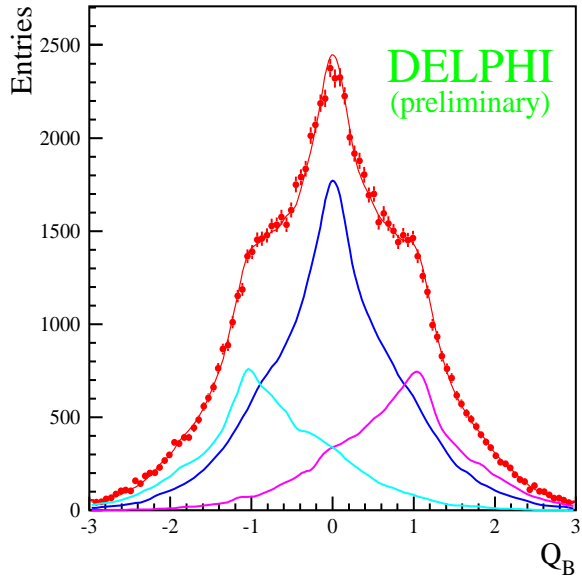,height=3in,width=3in}}
\caption{The DELPHI distribution of the net charge ($Q_b$) of particles
associated with a $b$-hadron
decay where $Q_b$ is determined by weighting the charge of particles by the probability
that they originate from a $b$-hadron (points). The results of the fit are superimposed and the
neutral and charged components are shown.
\label{fig:vtxchg}}
\end{figure}

The fragmentation fractions $f_{B^+}$, $f_{B^0}$ can be extracted from the net charge of
particles associated with the secondary vertex. A net charge of 0 is obtained from $B^0$,
$B_s$, and neutral $b$-baryons, while a net charge of $\pm 1$ is obtained from $B^\pm$ and
charged $b$-baryons. Delphi has determined $f_{B^+}$ and $f_{B^0}$ by fitting the vertex charge
distribution (Fig.~\ref{fig:vtxchg}) and estimating~\cite{delphifrag} the contributions from $B_s$
($f_{B_s} = 0.097\pm 0.013$) and $b$-baryons ($f_{\Lambda_b}=0.106^{+0.037}_{-0.027}$ of which
7\% are charged, 93\% neutral).
Alternatively, the data are fit assuming $f_{B^+}=f_{B^0}$ and $f_{\Lambda_b}$ is allowed to vary.
 
The $b$-quark fragmentation function results are summarized in Table~\ref{tab:bfrag}.
In addition to the new results described above, the particle data group~\cite{pdg}
averages and a published ALEPH measurement~\cite{bfragaleph} of $f_{\Lambda_b}$
not included in the particle data group average are given.
The new results are in good agreement with the particle data group averages and
substantially reduce the uncertainties in the $b$-quark fragmentation fractions.
\begin{table}[hbt]
\begin{center}
\caption{Summary of recent $b$-quark fragmentation fraction measurements.
The notation $f_{B^0}=f_{B^+}$ indicates a measurement of $f_{B^0}$, $f_{B^+}$
where they were constrained to be equal.}\label{tab:bfrag}
\vspace{0.2cm}
\begin{tabular}{|l|c|c|} 
\hline
\raisebox{0pt}[12pt][6pt]{Source} & 
\raisebox{0pt}[12pt][6pt]{Quantity} & 
\raisebox{0pt}[12pt][6pt]{Measurement} \\
\hline
\raisebox{0pt}[12pt][6pt]{CDF} &
\raisebox{0pt}[12pt][6pt]{$f_{B^0}$} &
\raisebox{0pt}[12pt][6pt]{$0.325\pm 0.048$} \\
\hline
\raisebox{0pt}[12pt][6pt]{DELPHI} &
\raisebox{0pt}[12pt][6pt]{$f_{B^0}$} &
\raisebox{0pt}[12pt][6pt]{$0.382^{+0.030}_{-0.038}$} \\
\hline
\hline
\raisebox{0pt}[12pt][6pt]{CDF} &
\raisebox{0pt}[12pt][6pt]{$f_{B^+}$} &
\raisebox{0pt}[12pt][6pt]{$0.422\pm 0.062$} \\
\hline
\raisebox{0pt}[12pt][6pt]{DELPHI} &
\raisebox{0pt}[12pt][6pt]{$f_{B^+}$} &
\raisebox{0pt}[12pt][6pt]{$0.415\pm 0.012$} \\
\hline
\hline
\raisebox{0pt}[12pt][6pt]{PDG (1998)} &
\raisebox{0pt}[12pt][6pt]{$f_{B^0} = f_{B^+}$} &
\raisebox{0pt}[12pt][6pt]{$0.397^{+0.018}_{-0.022}$} \\
\hline
\raisebox{0pt}[12pt][6pt]{CDF} &
\raisebox{0pt}[12pt][6pt]{$f_{B^0}=f_{B^+}$} &
\raisebox{0pt}[12pt][6pt]{$0.371\pm 0.024$} \\
\hline
\raisebox{0pt}[12pt][6pt]{DELPHI} &
\raisebox{0pt}[12pt][6pt]{$f_{B^0}=f_{B^+}$} &
\raisebox{0pt}[12pt][6pt]{$0.417\pm 0.013$} \\
\hline
\hline
\raisebox{0pt}[12pt][6pt]{PDG (1998)} &
\raisebox{0pt}[12pt][6pt]{$f_{B_s}$} &
\raisebox{0pt}[12pt][6pt]{$0.105^{+0.018}_{-0.017}$} \\
\hline
\raisebox{0pt}[12pt][6pt]{CDF} &
\raisebox{0pt}[12pt][6pt]{$f_{B_s}$} &
\raisebox{0pt}[12pt][6pt]{$0.162\pm 0.045$} \\
\hline
\raisebox{0pt}[12pt][6pt]{DELPHI} &
\raisebox{0pt}[12pt][6pt]{$f_{B_s}$} &
\raisebox{0pt}[12pt][6pt]{$0.088\pm 0.023$} \\
\hline
\hline
\raisebox{0pt}[12pt][6pt]{PDG (1998)} &
\raisebox{0pt}[12pt][6pt]{$f_{\Lambda_b}$} &
\raisebox{0pt}[12pt][6pt]{$0.101^{+0.039}_{-0.031}$} \\
\hline
\raisebox{0pt}[12pt][6pt]{ALEPH} &
\raisebox{0pt}[12pt][6pt]{$f_{\Lambda_b}$} &
\raisebox{0pt}[12pt][6pt]{$0.102\pm 0.028$} \\
\hline
\raisebox{0pt}[12pt][6pt]{CDF} &
\raisebox{0pt}[12pt][6pt]{$f_{\Lambda_b}$} &
\raisebox{0pt}[12pt][6pt]{$0.090\pm 0.030$} \\
\hline
\raisebox{0pt}[12pt][6pt]{DELPHI} &
\raisebox{0pt}[12pt][6pt]{$f_{\Lambda_b}$} &
\raisebox{0pt}[12pt][6pt]{$0.069\pm 0.030$} \\
\hline
\end{tabular}
\end{center}
\end{table}
\vspace*{3pt}
\section{Charmonium Production}
Charmonium production in high energy collisions probes the evolution of an
initial state consisting of one or more gluons to a color singlet final state.
In the color singlet model,\cite{colorsinglet} the color singlet final
state is produced directly, as in $gg\to \chi_c$ or $ggg\to J/\psi$,
and the dominant source of $J/\psi$ production
expected to be from $b$-hadron and $\chi_c\to\gamma J/\psi$ decay.
However, CDF~\cite{cdfjsi} and \D0~\cite{d0jsi} have measured $J/\psi$ production
cross sections significantly larger than predicted by the color singlet model
with a large fraction of the $J/\psi$ produced directly and not originating
in feeddown from $\chi_c$ or $b$-hadron decay.
Furthermore, CDF has measured the $\psi^\prime$ production cross section to be a
factor of $\sim 50$
larger than predicted by the color singlet model.

An alternative way to produce a color singlet final state is described by the
color octet model,\cite{coloroctet} where a color octet $c\bar c$
state is produced in the hard collision. The color octet state then
evolves to a color singlet state through the emission of soft gluons,
as in $gg\to c\bar c \to gJ/\psi$ or $g\to c\bar c \to ggJ/\psi$.
New measurements that test the octet model predictions were presented~\cite{jesik}
at the conference and are described below.

\begin{figure}[hbtp]
\centerline{\protect\psfig{figure=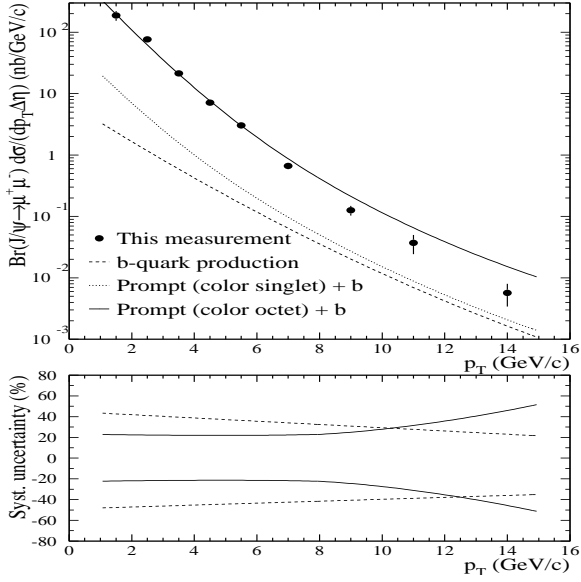,height=3in,width=3in}}
\caption{The $p_T$ dependence of the $J/\psi$ production cross section for
$2.5<|\eta_{J/\psi}|<3.7$ (upper figure). Shown are the \D0\ data with
statistical errors (points) and the prediction of the color octet model (solid line).
The lower figure shows the systematic error in the $J/\psi$ cross section (solid line)
and the uncertainty due to $J/\psi$ polarization (dashed line). \label{fig:fwdpt}}
\end{figure}

\begin{figure}[hbtp]
\centerline{\protect\psfig{figure=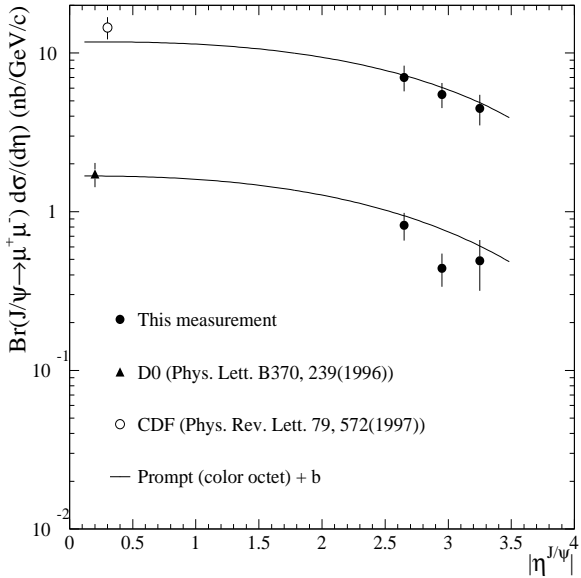,height=3in,width=3in}}
\caption{The pseudorapidity dependence of the $J/\psi$ production cross section
for $p_T>5$ GeV/$c$ (upper points) and $p_T>8$ GeV/$c$ (lower points). 
Shown are the \D0\ data (points) with statistical and systematic errors, but not
including the uncertainty due to the $J/\psi$ polarization. The curves
show the cross section predicted by the color octet model. \label{fig:fwdeta}}
\end{figure}

The non-perturbative matrix elements for $J/\psi$ production in the color octet model
were obtained by fitting the CDF $J/\psi$ $p_T$ spectrum in the central region
($|\eta_{J/\psi}|<0.6$).
Given these matrix elements, the $J/\psi$ cross section in other kinematic regions
is predicted by the color octet model.
\D0\ has studied $J/\psi$ production in the forward region ($2.5<|\eta_{J/\psi}|<3.7$)
using its small angle muon spectrometer.\cite{d0fwdjsi}
The dependence of the forward $J/\psi$ cross section on $p_T$ and $\eta_{J/\psi}$
is shown in Figs.~\ref{fig:fwdpt}--\ref{fig:fwdeta}.
The color octet model is reasonably successful in predicting the behavior of the
forward $J/\psi$ cross section with no further tuning of parameters.
 
\begin{figure}[hbtp]
\centerline{\protect\psfig{figure=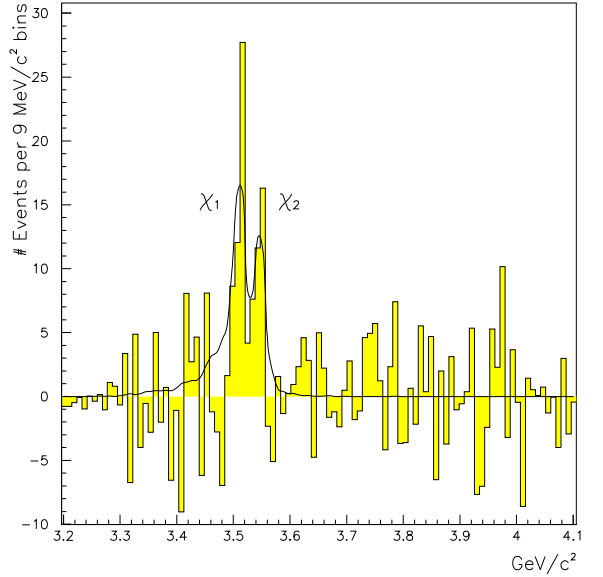,height=3in,width=3in}}
\caption{The E771 background subtracted $e^+e^-J/\psi$ invariant mass distribution.
The curve shows a fit to the $\chi_1$ and $\chi_2$ peaks. \label{fig:e771}}
\end{figure}

The production of $\chi_c$ mesons in proton--silicon collisions has been studied by 
E771 at Fermilab. Photon conversions are used to reconstruct $\chi_c\to \gamma J/\psi$
decays (see Fig.~\ref{fig:e771}).
The $\chi_1$ and $\chi_2$ states are resolved and the ratio of their production cross
sections was measured to be
\begin{equation}
\sigma(\chi_1)/\sigma(\chi_2) = 0.59\pm 0.19\ {\rm (stat)}\ \pm 0.07\ {\rm (syst).}
\label{eq:e771}
\end{equation}
When combined with other measurements of $\sigma(\chi_1)/\sigma(\chi_2)$ in proton--nucleon
collisions, $\sigma(\chi_1)/\sigma(\chi_2)=0.31\pm 0.14$ is obtained,\cite{karlapc}
in good agreement with the color octet prediction~\cite{chislac} of $\sim 0.3$.

The color octet model can also be tested by measuring the rate for prompt $J/\psi$
production in $Z$ boson decays.\cite{cho,cheung} The dominant production
mechanism is expected to be gluon fragmentation followed by soft gluon emission
to reach a color singlet state: $Z\to q\bar q g$, $g\to gg J/\psi$.
New results were presented at the conference on the observation of prompt $J/\psi$
mesons by L3 in $Z$ boson decays.\cite{jesik,l3jsi}

\begin{figure}[hbtp]
\centerline{\protect\psfig{figure=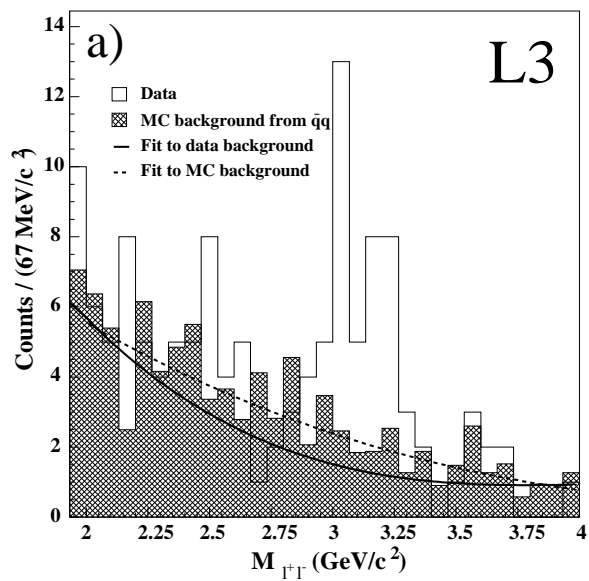,height=3in,width=3in}}
\caption{The L3 distribution of $\ell^+\ell^-$ invariant mass for events with minimal
additional energy in a 30$^\circ$ cone about the $J/\psi$ direction. The shaded 
area shows the expected background. \label{fig:l3a}}
\end{figure}

\begin{figure}[hbtp]
\centerline{\protect\psfig{figure=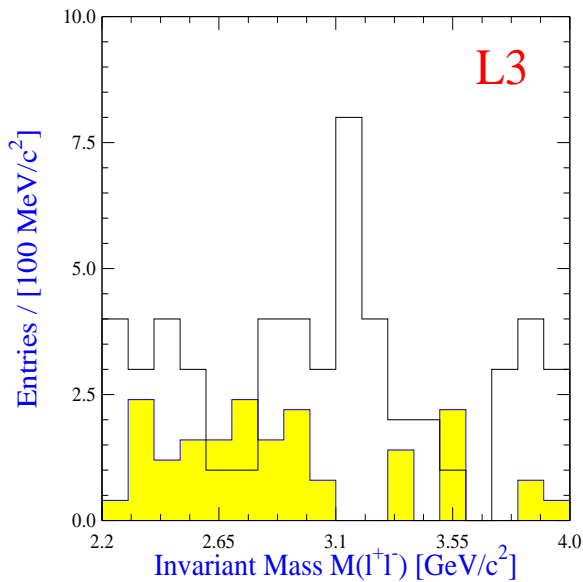,height=3in,width=3in}}
\caption{The L3 distribution of $\ell^+\ell^-$ invariant mass for events with no
jets within 40$^\circ$ of the $J/\psi$ direction. The shaded area shows the
expected background. \label{fig:l3b}}
\end{figure}

The decay of $b$-hadrons to $J/\psi$ is the dominant source of $J/\psi$ mesons observed
in $Z$ boson decay. The signature for prompt $J/\psi$ decay is that the $J/\psi$ is
isolated and not associated with a jet.
Two different isolation criteria are used: minimal additional energy in a 30$^\circ$ cone
about the $J/\psi$ direction (see Fig.~\ref{fig:l3a}) and no jets within 40$^\circ$ of the $J/\psi$
(see Fig.~\ref{fig:l3b}).
The fraction of $J/\psi$ decays from prompt production is calculated for both isolation criteria
and is found to be consistent.
Combining the results for the two criteria, the prompt $J/\psi$ branching ratio is found
to be
\begin{equation}
BR(Z\to J/\psi X) = (2.6\pm 0.7\pm 0.5\ ^{+0.5}_{-0.3})\times 10^{-4}
\label{eq:l3}
\end{equation}
where the first error is statistical, the second error is systematic, and the third error
is theoretical.
The measurement is in good agreement with the color octet prediction of
$2.8\times 10^{-4}$.

One discrepancy between data and the color octet model has been the $J/\psi$ photoproduction
cross section near $z=1$.\cite{h1data,heraprob}
However, it is argued~\cite{mehen} that the NLO calculation
breaks down in this region and requires resummation of higher order logarithmic terms.
Further tests~\cite{mehen} of the color octet model can be performed by measuring
the polarization of high $p_T$ $J/\psi$ mesons in hadron collider experiments and
leptoproduction of $J/\psi$ mesons at large $Q^2$.
\section*{Acknowledgements}
The author is grateful for the help and assistance provided by
the parallel session speakers:
Emanuela Barberis,
Mary Anne Cummings,
Ugo Gasparini,
Andre Hoang,
Rick Jesik,
Johann Kuehn,
Alan Litke,
Tom Mehen,
Lynne Orr,
Matthias Neubert,
Tomasz Skwarnicki,
Fr\'ed\'eric Stichelbaut,
Kirsten Tollefson,
Yorgos Tsipolitis, and
Weiming Yao.
The author would also like to thank Roy Briere, Bob Clare, Karla Hagan, and
Rob Roser, and Wendy Taylor for their assistance in providing material.
This work is supported in part by the U.S. Department of Energy 
Grant DE-FG02-91ER40688.
\section*{References}
\end{document}